\documentstyle[12pt,epsf,epsfig]{article}
\def\be{\begin{equation}}
\def\ee{\end{equation}}
\def\ba{\begin{eqnarray}}
\def\ea{\end{eqnarray}}
\def\bq{\begin{quote}}
\def\eq{\end{quote}}
\def\PL{{ \it Phys. Lett.} }
\def\PRL{{\it Phys. Rev. Lett.} }
\def\NP{{\it Nucl. Phys.} }
\def\PR{{\it Phys. Rev.} }

\begin{document}

\thispagestyle{empty}
\begin{flushright}
OUTP 97-38P\\
SU-ITP-97/44\\
UMN-TH-1609/97\\
hep-th/9711027\\
November 1997
\end{flushright}
\vspace*{1cm}
\begin{center}
{\Large \bf  Cos(M)ological Solutions in \\
M- and String Theory}
 \\
\vspace*{1cm}
Nemanja Kaloper\footnote{E-mail:
kaloper@hepvs6.physics.mcgill.ca}\\
\vspace*{0.2cm}
{\it Department of Physics, Stanford University}\\
{\it Stanford, CA 94305-4060, USA}\\
\vspace*{0.4cm}
Ian I. Kogan\footnote{E-mail: i.kogan1@physics.oxford.ac.uk}\\
\vspace*{0.2cm}
{\it Theoretical Physics, Department of Physics, Oxford University}\\
{\it 1 Keble Road, Oxford, OX1 3NP,  UK}\\
\vspace*{0.4cm}
Keith A. Olive\footnote{E-mail: olive@mnhep.hep.umn.edu}\\
\vspace*{0.2cm}
{\it School of Physics and Astronomy,
University of Minnesota}\\
{\it Minneapolis, MN 55455, USA}\\
\vspace{2cm}
ABSTRACT
\end{center}
We consider solutions to the cosmological equations of motion in
11 dimensions with and without 4-form charges. We show explicitly
the correspondence between some of these solutions and known solutions in
10 dimensional string gravity.  New solutions involving combinations of
4-form charges are explored.  We also speculate on the possibility of
removing curvature singularities present in 10D theories by oxidizing
to 11D.

\vfill
\setcounter{page}{0}
\setcounter{footnote}{0}
\newpage
\section{Introduction}

There is little doubt or discussion that classical
Friedmann-Robertson-Walker cosmology provides an
excellent description of the evolving
Universe at late times (say nucleosynthesis and beyond). Indeed, there is
little reason to doubt the validity of FRW back to very early times
corresponding to the GUT epoch.  At some point however, it is
reasonable to suppose that Einstein gravity is modified, and at present,
the only consistent modification available is due to string theory.
A strong argument in favor of such a modification is that it is not
possible to bring Einstein's General Relativity in full accord with
Quantum Mechanics, and hence, accepting the postulates of
Quantum Mechanics,
altering theory of gravity seems inevitable. In the regime of large
curvatures these alterations should be expected
to play a very significant role.
Since the singularity theorems of Hawking and Penrose \cite{he} state that
such large curvature regions are a generic state of the early Universe,
it then seems plausible to assume that in this epoch the effects of
quantum gravity strongly influence the evolution of the Universe.
There have been numerous efforts attempting to explore the effects
of string theory in cosmology \cite{dilinf} - \cite{tw}. One could
characterize the main aims of many of these studies as either an
attempt to utilize the additional degrees of freedom in the massless
sector to induce inflation \cite{dilinf,bd,other,nko1,tsey,CLO,gv} or
developing arguments on the type of modifications to Einstein gravity
which are necessary to avoid a cosmological singularity
\cite{wind,art,dp,bv,kmo1,kmo2,bm}.

Much of the emphasis in previous work on string cosmology has been
on the modifications to Einstein gravity \cite{act}. The modifications
were accommodated either by enlarging the
zero mass sector with the inclusion of the
dilaton and axion/moduli fields
or by considering higher curvature terms described by the low energy string
action. Dilaton/axion corrections do indeed have an effect on the
equations of motion at early times and lead to new cosmological solutions.
They are not however very conducive to
a deSitter, or a more general inflationary phase,
at least not
without resorting to supersymmetry breaking potentials
for trapping the dilaton \cite{bd}.
As an alternative, much effort has been invested
in trying to resolve the standard
cosmological problems with a Pre-Big-Bang phase \cite{gv},
which appears in the general space of solutions
to the string dilaton/gravity system.  It still remains to be seen,
whether or not such models can successfully solve all of the
problems normally associated with inflation and
produce density perturbations
consistent with the COBE measurements of
the microwave background anisotropy.

Another interest of string gravity is
the problem of an initial (or final) cosmological
singularity. The attempts to address it included the
use of winding modes wrapped around spatial directions \cite{wind},
higher-derivative/higher-genus induced corrections \cite{art},
decompactification to higher
dimensions with simultaneous insertion of D-brane type matter
sources \cite{dtype}, instanton-like constructions \cite{inst},
and models with generalized scalar-tensor couplings
\cite{Bar}-\cite{nko2}. Though we know that the simple types of
dilaton/axion  modifications to Einstein gravity considered up to now are
not  capable of removing singularities \cite{kmo1,kmo2,nko2},
progress has been made concerning the form of the corrections needed
\cite{bm}. Any such solution (at least from the 10D point of view) must
rely on non-perturbative features of the gravitational action
or come from some more complete theory
of gravity at the string scale. An effective field
theory approach which we
will return to later
was proposed by Damour and Polyakov \cite{dp}, where
the universality of the string coupling at the tree level
has been extended to all string loops.

Out of the morass of different weakly,
and strongly, coupled string theories,
M-theory is emerging as the single underlying theory capable of unifying
all particle interactions \cite{w,hw}. At this time, our understanding of
M-theory is still incomplete. While its various low energy limits, and
the links between them, are known, (which are the consistent string
theories and the 11D supergravity, related by the web of dualities),
the full description of the theory is still being sought for.
A candidate that has been proposed recently is the M(atrix) theory,
formulated as a large N-limit supersymmetric  matrix quantum mechanics
\cite{matrix}.
An interesting shift of the point of view that has emerged
out of these developments
is that the dilaton scalar field,
present in all known string theories, has been demoted
to merely another modulus field in the 11D supergravity.
This could have important consequences for cosmological applications.
The troubles with implementing conventional
inflationary scenarios in string theory arise because of the
dilaton and its couplings to the other modes in the string spectrum.
In the arena of 11D supergravity, such couplings are absent.
Some of the obstacles for inflation in
dilaton-plagued string theories could
perhaps be resolved by way of M-theory.
Let us be more specific:
the extreme weak and strong coupling limits
in string theory formulations correspond to the regimes where the
size of the eleventh dimension becomes very small or very large.
These limits sit in rather special portions of
the phase space of the full theory,
and perhaps should be viewed as unnatural.
Indeed, there seems to be no reason why
at a given very large energy scale
the dynamics should treat any direction
in the Universe any differently than
the others. On the other hand,
the present knowledge of the
low-energy limits of M-theory does
not seem to prefer one construction over the other.
Since the limits where the
moduli attain their extremes do exist
within reach of solutions in the phase space,
it then seems logical to see if they can be dynamically
understood from the M-theory point of view.
For example, we can imagine a scenario in string
theory where in the limit
when the moduli converge towards their extrema,
the effective stringy
description must be lifted to 11
dimensions, where some intrinsically
M-theoretic (and as yet unspecified)
mechanism saturates moduli
evolution. It might be possible to
find some inflationary scenario in this limit.
This scenario could work in string theories
essentially because of
duality, albeit the mechanism
might take a different guise there\footnote{In spirit,
this would be similar to the proposed (but to date unknown) scenario
for solving the cosmological constant problem in string theory. The
idea is to find a duality relationship between a string vacuum with
unbroken supersymmetries and a vacuum with all supersymmetries broken.
Then duality would guarantee that the cosmological constant, which is
zero in the vacuum with unbroken susy's must also vanish in the vacuum
without manifest susy's. Perhaps it is possible to hide inflation in string
theory in this sense too.}.

At this moment, we are far from being able to address
comprehensively the questions we pose above. However,
we can at least consider the known string
cosmologies from the (ad)vantage of the $11^{\rm th}$ dimension.
We will therefore concentrate here
on the oxidation of 10D type IIA superstring theory cosmologies
to 11D supergravity theory
with the geometric reinterpretation of the
string coupling. There have been several interesting
papers directly exploring cosmological
solutions based on the M-theory-inspired
action \cite{Lu} and making use of string dualities \cite{Luk}.
Furthermore, many of the earlier investigations of cosmological
models with higher-rank form-field charges in superstring
models \cite{form} can be directly incorporated into the framework
of the 11D supergravity
by dimensionally oxidizing the solutions. In this way,
one obtains a description of the known string cosmologies,
which treats the dilaton field on equal footing with the
other moduli fields. An immediate consequence
of this approach is that by means of $U$-duality one can flow between
different M-theory configuration, as exemplified in \cite{luov1}.
The possible advantages
of such a {\it modular democracy} remain to be investigated further.

Below we will survey several classes of cosmological solutions of
the 11D theory which can be reduced to
the solutions of string  dilaton gravity. We will
give the explicit relationship of the string and M-theory
solutions where applicable. We will
also study a case when the 4-form field strength carries two
different charges, a magnetic and an electric one, which do not
correspond to any of the combinations of form-field charges studied
in reduced models so far (but can be given a string theory
interpretation via Scherk-Schwarz dimensional reduction).
Most of the solutions still feature
the unattractive properties of their lower-dimensional stringy
relatives, in that they are singular and have running moduli,
and hence cannot be used for
building an inflationary scenario by themselves. However, we
will show a special example where a flat space in 11 dimensions, viewed by
an observer accelerated along one of the circles, reduces
to a singular dilatonic string cosmology. This example isn't really
inflation, but it moderates the singularity by dimensional
uplift. We will also
consider a possible M-theory interpretation of a curvature
singularity-free model based on the Damour-Polyakov universality
ans\"atz. In the stringy picture, this solution has a smooth metric,
but runs through the extremely strong coupling regime.
We will argue that this coupling singularity could be understood as the
process of decompactifying the radius of the eleventh dimension.
Finally, we will attempt to extend
the M-theory relation of the value of the
string coupling to the size of the $11^{\rm th}$
dimension, to a relation
between the renormalization group running of the coupling and the
cosmological evolution of this scale factor.

\section{String Theory Actions and Equations of Motions}

Before we begin our study of the general M-theory inspired,
11D supergravity action, it will be useful to review some of the
salient features of the effective field theory formulation of
string gravity as it pertains
to cosmological solutions. Neglecting for now the contribution to dynamics
from the 6D Calabi-Yau space, we can begin with the lowest order
4D effective action
of the NS-NS sector of any string construction \cite{act}:
\be
S = \int d^4x \sqrt{g} e^{-2 \phi} \Bigl\{ R
+ 4 (\nabla \phi)^2 - \frac{1}{12} H_{\mu \nu \lambda}
H^{\mu \nu \lambda} + 2\Lambda
\Bigr\}
\label{sact1}
\ee
The 3-form $H=dB$ is the field strength of the
Kalb-Ramond 2-form
$B_{\mu \nu}$. The stringy cosmological constant $\Lambda$ can
arise from central charge deficit in conformal field theory
constructions or by the reduction of higher rank form
fields, as we will see later.
In four dimensions, this 3-form is
dynamically dual to a
pseudoscalar axion field. The correspondence
is given by
\be
H_{\mu \nu \lambda}=\sqrt{2} e^{2 \phi} \sqrt{g}
   \epsilon_{\mu \nu \lambda \rho} \partial^{\rho} \chi
\ee
resulting in the replacement of the 3-form kinetic term
coupled to the inverse string coupling $e^{-2 \phi}$
by the pseudoscalar kinetic term
coupled to the string coupling itself. By means of a
simple conformal rescaling,
$g_{\mu\nu} \rightarrow e^{2\phi}g_{\mu\nu}$, the action can be put into
the Einstein frame, where the Planck mass is
constant (for simplicity, we set
$2 \kappa^2 =1$):
\be
S = \int d^4x \sqrt{\bar g} \Bigl\{ {\bar R}
- 2 (\nabla \phi)^2 - 2 e^{4 \phi} (\nabla \chi)^2 + 2 \Lambda e^{2\phi}
\Bigr\}
\label{sact1e}
\ee

With the FRW spatially flat ans\"atz for the metric,
\be
ds^2 = - n^2(t) dt^2 + a^2(t) d\vec x ^2
\ee
where $n$ is a gauge parameter (lapse function),
we can easily derive the equations of motion
from the action (\ref{sact1}) (see e.g. \cite{kmo2}). They
are
\begin{eqnarray}     \label{ee1}
&&\dot h = 2 h \dot \phi - 3 h^2 + \rho \nonumber \\
&& \dot \rho + 6 h \rho = 0 \\
&& 2 {\dot \phi}^2 + 3 h^2 - 6 h \dot \phi -
\rho / 2  + \Lambda = 0 \nonumber
\end{eqnarray}
where $h=\dot a/a$ is the string-frame Hubble parameter,
and $\rho=e^{4 \phi} {\dot \chi}^2$ is the effective energy density
of the pseudo-scalar axion field, and in the gauge $n=1$.

These equations of motion are straightforward
to solve. Many authors have already
considered various aspects of the solutions,
both in the Einstein frame and in the
string frame. The simplest case is certainly
the pure metric-dilaton solution
with vanishing cosmological constant
\cite{lindil1,lindil2,nko1,tsey,gv,gp}, which is given by
two classes of solutions separated by the curvature singularity,
used in \cite{gv} to construct the Pre-Big-Bang scenario:
\ba
\label{metdil}
ds_{+}^2 &=& -dt^2 + a_0^2 |\frac{t}{t_0}|^{\mp
\frac{2}{\sqrt{3}}} d\vec x^2 ~~~~~~~~~~~~~
e^{-2\phi} = e^{-2\phi_0} |\frac{t}{t_0}|^{1\pm\sqrt{3}}
~~~~~~~~~~ (t<0) \nonumber \\
ds_{-}^2 &=& -dt^2 + a_0^2 (\frac{t}{t_0})^{\pm \frac{2}{\sqrt{3}}}
d\vec x^2 ~~~~~~~~~~~~~
e^{-2\phi} = e^{-2\phi_0} (\frac{t}{t_0})^{1\mp\sqrt{3}}
~~~~~~~~~~ (t>0)
\ea
In the string frame, both branches
consist of two classes of solutions:
expanding and contracting.
The solutions for $t<0$ are by now widely referred to as the $(+)$ branch,
and those for $t>0$ as the $(-)$ branch. The proper definition
of branches  is derived from solving the quadratic
constraint equation in (\ref{ee1}) for $\dot \phi$. The sign of each branch
is determined by the sign of the square root which arises in
the solution, where the discriminant is not zero. If the discriminant
vanishes anywhere on the phase space trajectory, the branches
connect there. The solutions
are isotropic $T$-duals of each other.
A goal of the Pre-Big-Bang scenario is the connection of the expanding
solutions in the two branches, in such a way that the $(+)$-branch
chronologically precedes the
$(-)$-branch and hence the singularity would be removed. In spite
of some recent results \cite{bm}, it still remains to be seen if a
coherent and fully consistent description of branch-changing can be
found.  It is interesting to note that in the E
frame, both the expanding and contracting metrics degenerate to a single
Einstein frame metric, and that the only difference between the
two subclasses of solutions is the sign of the dilaton field. Since
in the Einstein frame the switch of the sign of the dilaton corresponds
to the classical form of the $S$-duality map, these solutions are also
$S$-duals of each other.

Further generalizations of these solutions can be easily obtained
with the help of generating techniques. For example, if we add
the cosmological constant, we can find new solutions starting
from (\ref{metdil}) and applying a solution-generating technique
described in \cite{nkredef}. Moreover, we can obtain solutions
with the axion field if we apply an $SL(2,R)$ duality rotation to
(\ref{metdil}), as described in \cite{STW,clw}. These solutions
can be written in terms of the functions $a$ and $\exp(-2\phi)$
describing the axion-less case. Following Copeland et al \cite{clw},
they are
\ba
\label{axion4d}
&& ~~~~~~~~ d\bar s^2 = (s^2 + r^2 e^{-4 \phi})
(-dt^2 + a^2(t) d\vec x^2) \nonumber \\
&&e^{2\bar \phi} = s^2 e^{2\phi} + r^2 e^{-2\phi} ~~~~~~~~~~~
\bar \chi = \frac{qs + pr e^{-4\phi}}{s^2 + r^2 e^{-4\phi}}
\ea
where $p$, $q$, $r$ and $s$ are real numbers satisfying $ps - qr =1$.
The 3-form axion field is $H = Q_A d^3\vec x$, in form notation,
where the constant charge is determined by the integration
constants. The solutions can be readily generalized to include
additional scalar moduli, which arise from the reduction of the
10D string theories.
Note that for these solutions each branch now contains only one
congruence of the system trajectories. This is because as the
string frame scale factor approaches zero, the dominant source
in the equations of motion is the axion, since its contribution
to the total stress-energy goes as $1/a^6(t)$. This term
then forces the Universe to bounce away from zero volume and start
expanding again. Hence the qualitative picture of evolution
in both branches is that the Universe begins in a stage of
contraction, reaches its minimal volume and starts expanding again
to infinity.  One should note however, that such an axion-driven bounce
does not allow one to evade the cosmological singularity.  Perhaps the
easiest way to see this, is to note that the bounce occurs at some small
but finite value of the scale factor and some large but still finite
value of the coupling.  When it occurs, the bounce changes only the sign
of the Hubble parameter and not the sign of the $\dot \phi$ and therefore
the coupling continues to grow.  The curvature singularities reside in
the regime of very large coupling which therefore can still be attained
in the axionic cosmologies. Since the Universe is now expanding, the
axion's contribution to $\rho$ is red-shifted away, and eventually the
Universe becomes dilaton dominated and therefore must inevitably run away
towards the singularity.  In other words, the bounce occurs at finite but
negative $t$ and still evolves towards the singularity at $t=0$.

Finally, we can
obtain solutions with spatial curvature, either by directly
solving differential equations for models with spatially
curved sections \cite{tsey,gp,tw} or
by using a Wick rotation and a
dimensional reduction of the 5D Schwarzschild black hole solutions
\cite{mology}. The equations of motion including spatial curvature
(but for simplicity excluding the stringy cosmological term and
the axion contribution) are
\be
\label{eomcurv}
\dot h = 2 h \dot \phi - 3 h^2 -2 \frac{k}{a^2}
~~~~~~ 2 {\dot \phi}^2 + 3 h^2 - 6 h \dot \phi + 3\frac{k}{a^2} =0
\ee
and the generic solutions are given \cite{tsey,mology,tw}
\ba
\label{curvsols}
&&ds^2 = \mu ({\cal C}^{1\pm\sqrt{3}}(\vartheta+\vartheta_0))
({\cal S}^{1\mp\sqrt{3}}(|\vartheta+\vartheta_0|))
\Bigl(-d\vartheta^2 + d\Omega_k\Bigl) \nonumber\\
&&~~~~~~~~e^{\phi} = \Bigl(\frac{{\cal C}(\vartheta+\vartheta_0)}
{{\cal S}(|\vartheta+\vartheta_0|)}\Bigr)^{\pm\sqrt{3}}
\ea
where $d \Omega_k$ is the metric on the maximally symmetric
3D spaces with constant curvature $k$, and ${\cal C}$ and
${\cal S}$ are the trigonometric or hyperbolic cosine and sine,
depending on whether $k=1$ or $k=-1$, respectively.
The parameters $\mu$ and $\vartheta_0$ are integration constants.
We should mention here that there also exist special
solutions for $k=1$ cases, when the string frame scale factor
depends linearly on the comoving time \cite{lindil2}.
Note here that in contrast to the spatially flat models,
the curved ones do not have an infinite amount of time
available for pole expansion. Rather, the closed $k=1$
solutions emerge out of the spatial curvature-controlled
singularity ($(+)$ branch) or end up in it ($(-)$ branch), while
the open $k=-1$ solutions begin in a contracting phase, and only
rebound later, pole-expanding for a finite amount of time
before hitting the curvature singularity. This has been used recently
to argue that Pre-Big-Bang viewed as inflation suffers from a fine
tuning problem \cite{tw}.

As we have seen above, the curvature singularity which separates
the $(+)$ and $(-)$ branches shows that near it the
cosmological evolution is dominated by the dilaton field
\cite{nko2}, where the string coupling $\lambda_s=\exp(\phi)$
diverges, and all other degrees of freedom become irrelevant.
Hence, all the solutions in this regime are extremely well
approximated by the pure metric-dilaton configuration.
Recently attempts have been made to dampen this singularity
with the higher derivative and/or higher genus contributions
to the equations of motion \cite{art,bm,venetal,sjrey}.
However, it has also
been noticed \cite{kmo1} that there exist solutions
in the model proposed by
Damour and Polyakov \cite{dp} where the effective coupling function
also diverges, but the strong coupling limit
string metric remains completely smooth.
In this context, the Damour-Polyakov universality ans\"atz amounts to
replacing the factor $e^{-2\phi}$ by a function $B(\phi) = e^{-2\phi}
+ c_0 + c_1 e^{2\phi} + ...$ in Eq. (\ref{sact1}). An action of this
form was considered in \cite{kmo1} in an attempt to achieve a
graceful exit from a pre-big bang phase in the dilaton-gravity cosmological
evolution. Such a solution was indeed found, however, with the
unpleasant aspect that though there were no space-time singularities
in the solution, there was a point in the evolution in which the function
$B(\phi)$ changes sign corresponding to a signature change in the metric.
Below we will speculate on an M-Theory interpretation of this type
of evolution.

From another perspective, extensions of the standard cosmological
solutions to higher dimensions have also been considered \cite{cd,many}.
There is a wide variety of motivations for such considerations which
we will not attempt to review here.  Most of them (in a cosmological
setting) are based on an ans\"atz for the metric of the form
\be
\label{tenprod}
ds^2 = - dt^2 + a^2(t) g_{ij} dx^i dx^j + b^2(t) g_{mn} dx^m dx^n
\ee
where $g_{ij}$ is assumed to be a maximally symmetric 3-space
and $g_{mn}$ some other metric describing the $d$ compactifed dimensions.
Considerable effort was expended to investigate the possibilities
that such a system could account for inflation, whereby the FRW portion of the
metric expands exponentially (or fast enough) at the expense of the
remaining $d$ dimensions.  Also, if we assume that the original
theory which we want to study (\ref{tenprod}) is devoid of the dilaton field,
we could retrieve the dilaton in lower dimensions.
Upon compactification, it can be seen that
the moduli from higher dimensions can
play the role of the dilaton, which was hoped to be
identified with the inflaton.
An approach of this kind, which most closely resembles the system we will
study below, is that of \cite{gss} based on 10D supergravity with the
action Eq. (\ref{sact1e}).
If we assume $g_{mn}$ to be maximally symmetric as well, and work in the
Einstein frame, where the effective dilaton field has the canonical kinetic term,
the correct cosmological equations of motion with the dilaton as
the only matter field can be simply
written as
\begin{eqnarray}
\label{gss1}
&&3{{\ddot a} \over a} + d {{\ddot b} \over b} = -2\dot \phi^2 \\
\label{gss2}
&&{{\ddot a} \over a} + 2 {{\dot a}^2 \over a^2} + {2k \over a^2} +
d {{\dot a} \over a} {{\dot b} \over b} = 0 \\
\label{gss3}
&&{{\ddot b} \over b} + (d-1) {{\dot b}^2 \over b^2} + {(d-1)k_d \over b^2} +
3 {{\dot a} \over a} {{\dot b} \over b} = 0 \\
\label{gss4}
&&\ddot \phi + (3 \frac{\dot a}{a} + d \frac{\dot b}{b} )\dot \phi = 0
\end{eqnarray}
A more detailed inspection of these equations
shows that regardless of the frame, the solutions for the scale factors
behave as powers of the comoving time
\cite{clw,clidw}. In addition, they are all singular, and
hence can still be grouped into different branches, much like when the
internal space is constant.
Thus, the system
(\ref{gss1})-(\ref{gss4}) does not admit conventional de-Sitter
inflationary solutions. (The de-Sitter solution in \cite{gss} stems from
a simple sign mistake in (\ref{gss1}).)

\section{Trans-dimensional Cos(M)ology}

As we have indicated above,
our main goal in this paper is to examine the cosmological implications
of the oxidation of 10D string theory to 11D M-theory in which we
will interpret the string coupling $e^{-2\phi}$ as the scale factor of
the $11^{\rm th}$ dimension \cite{w,hw}.
Our starting point therefore, will be the 11D supergravity action
\be
\label{act}
S=\int d^{11}x \sqrt{g} \Bigl\{R
- \frac{1}{48} F^2_{\mu_1 ... \mu_4} - \frac{1}{(4!)^2 (3!)^2}
\frac{\epsilon^{\mu_1 ... \mu_3 \nu_1 ... \nu_4 \lambda_1 ... \lambda_4}}
{\sqrt{g}} A_{\mu_1 ... \mu_3} F_{\nu_1 ... \nu_4}
F_{\lambda_1 ... \lambda_4} \Bigr\}
\ee
where  $R$ is the scalar curvature of
the 11D metric, and $A_{\mu_1 ... \mu_3}$ and
$F_{\nu_1 ... \nu_4} = 4 \partial_{[\nu_1} A_{\nu_2 ... \nu_4]}$
are the $3$-form potential and its $4$-form field strength. The last
term is the Chern-Simons term for $A$. Our conventions are
$g_{\mu\nu} = {\rm diag}(-1, \vec 1_{10})$,
$R^{\mu}{}_{\nu\lambda\sigma} = \partial_\lambda
\Gamma^{\mu}_{\nu\sigma} - ...$, $A = (1/3!) A_{\mu_1 ... \mu_3}
dx^{\mu_1} \wedge ... \wedge dx^{\mu_3}$ and
$F = d A = (1/4!) F_{\mu_1 ... \mu_4}
dx^{\mu_1} \wedge ... \wedge dx^{\mu_4}$.
We choose units such that in the
E frame we have $16\pi G_N=1$.

Before we investigate the equations of motion coming from the
action (\ref{act}), let us first reduce to 10D to further clarify our notation
which will follow closely that of Witten \cite{w}. Assuming that the
$11^{th}$ direction is compact, we can carry out
Kaluza-Klein reduction of (\ref{act}) to find
\begin{eqnarray}
\label{act10}
S & = & \int d^{10}x \sqrt{g_{10}} {\cal R}_{11} \Bigl\{R_{10} -
{\cal R}_{11}^2 {1 \over 4} F_{KK}^2 - \frac{1}{48} \bar F^2_{\mu_1 ...
\mu_4}  - \frac{1}{12} {\cal R}_{11}^{-2}
H^2_{\mu\nu\lambda} \nonumber \\
&& -
\frac{1}{384}
\frac{\epsilon^{\mu_1 \mu_2 \nu_1 ... \nu_4 \lambda_1 ... \lambda_4}}
{\sqrt{g}} B_{\mu_1 \mu_2} F_{\nu_1 ... \nu_4}
F_{\lambda_1 ... \lambda_4} \Bigr\}
\end{eqnarray}
where ${F_{KK}}_{\mu\nu} = 2 \partial_{[\mu} V^{11}_{\nu]} $ is the
field strength of the Kaluza-Klein gauge field coming from the metric
and the reduced
$2$-form is
$B_{\mu\nu} = A_{\mu\nu 11}$, and its
$3$-form field strength is $H_{\mu\nu\lambda} = \nabla_{\mu} B_{\nu\lambda}
+ {\it cyclic ~ permutations}$.
The reduced $4$-form field strength $\bar F$ acquires Chern-Simons type
couplings to the reduced $2$- and $1$-forms:
$\bar F_{\mu\nu\lambda\sigma} = F_{\mu\nu\lambda\sigma} + (A_{\mu} H_{\nu\lambda\sigma}$
$+ ~cyclic ~permutations)$.
After a conformal rescaling $g_{10} = {\cal R}_{11}^{-1} g_s$, and defining the
dilaton by $\exp(2\phi/3) = {\cal R}_{11}$, we find
\begin{eqnarray}
\label{act10s}
S &=& \int d^{10}x \sqrt{g_{s}} \Bigl\{e^{-2\phi} \Bigl(R_{s}
+ 4(\nabla \phi)^2 - {1 \over 12} H^2 \Bigr) \nonumber \\
&& - \frac{1}{48} \bar F^2_{\mu_1 ... \mu_4} - \frac{1}{4} F_{KK}^2
- \frac{1}{384} \frac{\epsilon^{\mu_1 \mu_2 \nu_1 ... \nu_4
\lambda_1 ... \lambda_4}}
{\sqrt{g}} B_{\mu_1 \mu_2} F_{\nu_1 ... \nu_4}
F_{\lambda_1 ... \lambda_4} \Bigr\}
\end{eqnarray}
This is precisely the effective action which describes the low
energy limit of the IIA superstring. We can recognize the first
three terms as the NS-NS sector of the theory, and the remaining ones
as the RR sector.
It is easy to rewrite this action in
the ten-dimensional Einstein frame,
by a further conformal rescaling $g_s = e^{\phi/2} g_E$.
The action (\ref{act10s}) can be reduced further to make contact with type IIB and
heterotic theories.

Since we want to relate the M-theory cosmological solutions
to the stringy cosmologies studied so far, we will
assume that the base manifold is split into
\be
\label{topology}
{\cal M}_{11} = {\cal R}_t \times {\cal M}_3^{k=0} \times
{\cal S}^1 \times {\cal M}_6^{CY}
\ee
where ${\cal R}_t \times {\cal M}_3^{k=0}$ is the spatially flat
$4D$ FRW Universe, ${\cal S}^1$ is a circle corresponding to the
$11^{th}$ dimension, and ${\cal M}^{CY}_6$
is some Calabi-Yau manifold, whose specifics are not necessary for
our purposes here. Here we will ignore all graviphotons which could
arise from a generic dimensional reduction of the metric.
First, such degrees of freedom cannot arise from
mixing the Calabi-Yau sector with the space-time, since the
topology of Calabi-Yau spaces does not support harmonic 1-forms,
that would be needed to carry the reduced gauge fields.
Furthermore, while we in principle can obtain an $U(1)$
field from the metric, which describes D0-branes
in the resulting type IIA theory, at this time
we wish to consider only gauge-neutral cosmological solutions.
This is because our focus here is on making the
M-theory reinterpretation of the known string cosmological solutions.
Indeed, such setting of the gauge field to zero
is consistent with the equations of motion, which
are homogeneous in the cross-terms.
The ans\"atz for the metric is then
\be
\label{metbackground}
ds^2 = - n^2(t) dt^2 + a^2(t) d\vec x ^2 + b^2(t) G^{CY}_{MN}(y) dy^M dy^N
+ {\cal R}_{11}^2(t) d \varphi^2
\ee
where we again retain the lapse function $n^2(t)$ since we will work in the
action. The functions $a(t)$, $b(t)$ and ${\cal R}_{11}(t)$ are the radii
of the three subspaces ${\cal M}_3^{k=0}$, ${\cal M}^{CY}_6$ and
$S^1$ of (\ref{topology}), respectively,
from the point of view of the $11D$ observer.
After the reduction to 10D, ${\cal R}_{11}$ is related to the
string coupling constant.  $G^{CY}_{MN}(y)$ in (\ref{metbackground}) is
the metric on the Calabi-Yau $3$-fold, which depends
only on the Calabi-Yau coordinates $y^M$.
We will not need the explicit dependence
of $G^{CY}$ on $y$ in this work.
It will be sufficient to keep in mind that
this metric is Ricci-flat, i.e. that
${\cal R}_{MN} = 0$.
Still, this subspace influences the overall
dynamics because it is nontrivially warped in $11D$, via the
scale factor $b(t)$.

We will also ignore the possibility
that the Calabi-Yau factor in (\ref{topology}) can have nontrivial
harmonic two-forms which can support additional reduced  U(1)  gauge
fields. Here $F$ will be completely supported by the remainder
of the base manifold, i.e. the ${\cal R}_t \times {\cal M}_3^{k=0} \times
{\cal S}^1$ subspace.
The Bianchi identity for $F$ is $dF = 0$, which follows from
the definition $F=dA$, and in terms of the components, becomes
$\partial_{[\rho} F_{\mu\nu\lambda\sigma]} = 0$.
By varying the action (\ref{act}) with respect to
$A_{\nu\lambda\sigma}$ we find the propagation equation for $F$.
The gauge dynamics is therefore determined by
\ba
\label{Feoms}
&&\nabla_\mu F^{\mu\nu\lambda\sigma} = \frac{1}{32 (3!)^2}
\frac{\epsilon^{\nu\lambda\sigma\mu_1 ... \mu_4 \rho_1 ... \rho_4}}
{\sqrt{g}} F_{\mu_1 ... \mu_4} F_{\rho_1 ... \rho_4} \nonumber \\
&&\partial_{[\rho} F_{\mu\nu\lambda\sigma]} = 0
\ea
If we restrict our attention to those configurations where
$F$ lives only in the ${\cal R}_t \times {\cal M}_3^{k=0} \times
{\cal S}^1$ subspace, and require that the three-space ${\cal M}_3^{k=0}$
is isotropic, we see that $F$ must be proportional to an exterior
product of the volume form on ${\cal M}_3^{k=0}$ and a one-form.
The two linearly independent possibilities for this one-form
are $d\varphi$ and $dt$. Hence we are left with
$F_{0ijk}$ and $F_{\varphi ijk}$.  It is quite clear that one can not
consider a case of non-zero
$F_{0\varphi jk}$,  because of the symmetry of the three-space
${\cal M}_3^{k=0}$. There are no invariant antisymmetric tensors of
rank 2 (as well as rank $1$) in ${\cal M}_3^{k=0}$ and because of
this any non-zero value of  $F_{0\varphi jk}$ will destroy the symmetry
of the three-dimensional space.
If we take into account the 6D Calabi-Yau indices,
we can write down objects like $F_{0\varphi AB}$ where
$A,B$ are in the Calabi-Yau subspace,
which has harmonic two-forms and hence
admit such terms. However, from the point of view of 4 dimensions,
terms of this type behave just like scalar fields,
after we reduce on $\varphi$.
Their contributions are in principle
consistent with the presence of
a maximally symmetric subspace in
four dimensions, and there is no {\it a priori}
reason to rule out such terms.
Nevertheless, here we will for simplicity set
these terms equal to zero.

With this, we see that in form notation, we can write down the
$4$-form as follows:
\be
\label{formans}
F = F_{0123} dt \wedge \Omega_3 + F_{\varphi 123} d\varphi \wedge \Omega_3
\ee
Here $\Omega_3 = d^3 \vec x$ is the comoving volume form of the three-space
${\cal M}_3^{k=0}$. Since the
Chern-Simons source in the first of Eq. (\ref{Feoms}) is proportional
to $F \wedge F$, it is always zero for the
backgrounds we consider, and hence we will ignore it from now on
\cite{Luk}.

Let us now solve the equations for $F$. For $F_{0ijk}$, since we
assume that it depends only on $t$, the Bianchi identity is
vacuous, and so are the equations
$\nabla_\mu F^{\mu 0jk} = 0$. The remaining Euler-Lagrange equation yields
\be
\label{EL1}
\nabla_{\mu} F^{\mu ijk} = \frac{1}{\sqrt{g}} \partial_\mu
\Bigl(\sqrt{g} F^{\mu ijk} \Bigr) =
- \frac{1}{a^3 b^6 {\cal R}_{11}} \frac{d}{dt}
\Bigl(\frac{b^6 {\cal R}_{11}}{a^3} F_{0ijk} \Bigr) = 0
\ee
and so,
\be
\label{Fsol1}
F_{0ijk} = P \frac{a^3}{b^6 {\cal R}_{11}} \epsilon_{0ijk}
\ee
where $P$ is a constant of integration and $\epsilon_{0ijk}$
is the $4D$ Levi-Civita symbol. We recognize this as the monopole ans\"atz
of Freund and Rubin \cite{fr}, and also Englert \cite{e}.
Similar cosmological backgrounds were considered
in \cite{many}. We can now look at the other mode, $F_{\varphi ijk}$.
The Euler-Lagrange equations in (\ref{Feoms})
are trivial, since they contain only derivatives with respect to $\varphi$ and
$x^k$. However, the Bianchi identity for this mode gives
\be
\label{B2}
\dot F_{\varphi ijk} dt \wedge d\varphi \wedge \Omega_3 = 0
\ee
and hence
\be
\label{Fsol2}
F_{\varphi 123} = Q
\ee
where $Q$ is another integration constant. In terms of the
potential $A_{ijk}$, we can write this mode as
$A = Q \varphi \Omega_3$. The linear dependence of the potential
on the compact
coordinate means that this solution is the Scherk-Schwarz
mode of $A$, which corresponds to a
4-brane wrapped around the circle $\varphi$. These generalized
reductions were considered as means of breaking supersymmetry
\cite{scsc}, and in contemporary developments have found the
interpretation of $p$-branes wrapped around longitudinal
tori \cite{cowd}.
The combined solution therefore is
\be
\label{Fsol}
F = \frac{P}{b^6 {\cal R}_{11}} a^3 dt \wedge \Omega_3 + Q d\varphi \wedge \Omega_3
\ee

The Einstein equations of motion which are obtained by the
variation of the action (\ref{act}) with respect to the $11D$ metric
$g_{\mu\nu}$ are
\be
\label{Eeqs}
R_{\mu\nu} - \frac{1}{2} g_{\mu\nu} R = \frac{1}{12}
F_{\mu\lambda\sigma\rho} F_{\nu}{}^{\lambda\sigma\rho} -
\frac{1}{96} g_{\mu\nu} F^2_{\lambda\sigma\rho\omega}
\ee
because the Chern-Simons term in (\ref{act}) does
not depend on the metric, and hence does not contribute
to the stress-energy tensor of $F$. However, it is much simpler
to work in the action, since the background (\ref{metbackground}),
(\ref{Fsol}) depends nontrivially only on time $t$.
We therefore dimensionally reduce
the $11D$ action (\ref{act}) to a $1D$ one, and then vary
it with respect to the independent degrees of freedom
$a$, $b$ and ${\cal R}_{11}$, and the lapse $n$.
We can ignore the Chern-Simons term, since it doesn't
contribute to either the equations of motion for $F$
(by virtue of (\ref{formans})) or the gravitational equations
of motion, as we see from (\ref{Eeqs}). To proceed, we first need
the Ricci scalar of $g_{\mu\nu}$. The easiest way to find it
is to use the tangent space representation, which
is given in terms of the $11$-bein
\be
\label{tangsp}
ds^2 = \eta_{ab} e^a e^b  ~~~~~~~
e^0 = n d t ~~~~~~~ e^k = a(t) dx^k ~~~~~~~ e^M = b(t) E^M
~~~~~~~ e^\varphi = {\cal R}_{11} d\varphi
\ee
and $E^M$ are the internal $6$-bein of the Calabi-Yau
$3$-fold ${\cal M}^{CY}$, such that
$G^{CY}_{MN} = \delta_{KL} E^K{}_M E^L{}_N$.
The capital latin indices run from $1$ to $6$ and
$\delta_{KL}$ is just the $6\times 6$ unit matrix. This
``bastard" split of the $11$-bein is similar to what is used
in the studies of the more complicated non-diagonal Bianchi models \cite{rysh}.
The next step is to determine the spin connexion $1$-forms $\omega_{ab}$.
Since we assume that the background (\ref{metbackground}) is
torsion-less, we can use $d e^a = - \omega^a{}_b \wedge e^b$ to find
the spin connexion.
This gives $\omega^k{}_0 = \frac{a'}{n a} e^k$,
$\omega^M{}_0 = \frac{b'}{nb} e^M$,
$\omega^\varphi{}_0 = \frac{{\cal R}_{11}'}{n{\cal R}_{11}} e^\varphi$
where the prime denotes derivatives with respect to time.
Defining the internal Calabi-Yau spin connexion
$\zeta_{MN}$ (details of which are not
necessary for our purposes),
we find that $\omega^M{}_N = \zeta^M{}_N$. The set of connexion
$1$-forms with all indices lowered for convenience is simply
\be
\label{connex}
\omega_{k0} = \frac{a'}{n a} e^k ~~~~~~~
\omega_{M0} = \frac{b'}{nb} e^M ~~~~~~~
\omega_{MN} = \zeta_{MN} ~~~~~~~
\omega_{\varphi 0} = \frac{{\cal R}_{11}'}{n{\cal R}_{11}} e^\varphi
\ee
The next step is to work out the curvature $2$-forms, using
$R_{ab} = d \omega_{ab} + \omega_{ac} \wedge \omega^{c}{}_{b}$.
We recall here that in the Calabi-Yau sector, we will
have the intrinsic
curvature ${\cal R}_{KL} = (1/2) {\cal R}_{KLMN} E^M \wedge E^N$.
Using the curvature forms,
we can obtain the curvature components in the tangent basis.
The tangent space curvature components
are (normalized by $R_{ab} = (1/2) R_{abcd} e^c \wedge e^d$)
\ba
\label{curvcomp}
&&R_{k0j0} = -(\frac{1}{n} (\frac{a'}{na})' + \frac{a'^2}{n^2 a^2})
\delta_{kj} ~~~~~~~ R_{M0L0} = -(\frac{1}{n} (\frac{b'}{nb})'
+ \frac{b'^2}{n^2 b^2}) \delta_{ML} \nonumber \\
&&R_{\varphi 0 \varphi 0} = -(\frac{1}{n} (\frac{{\cal R}_{11}'}{n{\cal R}_{11}})'
+ \frac{{\cal R}_{11}'^2}{n^2 {\cal R}_{11}^2})
~~~~~~~ R_{\varphi k \varphi j} = \frac{{\cal R}_{11}'a'}{n^2{\cal R}_{11}a}
\delta_{kj} \nonumber \\
&&R_{\varphi M \varphi L} = \frac{{\cal R}_{11}'b'}{n^2{\cal R}_{11}b} \delta_{ML}
~~~~~~~ R_{jklm} = \frac{a'^2}{n^2a^2} (\delta_{jl} \delta_{km}
- \delta_{jm} \delta_{kl}) \\
&&R_{kMjL} = \frac{a'b'}{n^2ab} \delta_{kj} \delta_{ML} \nonumber \\
&&R_{MNLK} = \frac{1}{b^2}{\cal R}_{MNLK} + \frac{b'^2}{n^2 b^2}
(\delta_{ML} \delta_{NK} - \delta_{MK} \delta_{NL}))
\nonumber
\ea
The contraction of indices is the interior product, and
hence a tensor operation, so it does not depend on the basis. Hence
we can contract the indices of these components, using the flat
tangent space metric $\eta_{ab}$, to get the tangent space
Ricci tensor components: $R_{ab} = \eta^{cd} R_{acbd}$.
Contracting again, we get the Ricci scalar,
which of course is basis independent:
$R = \eta^{ab} \eta^{cd} R_{abcd}$.
The tangent space Ricci tensor is
\ba
\label{riccit}
&&R_{00} = -3(\frac{1}{n} (\frac{a'}{na})' + \frac{a'^2}{n^2 a^2})
-(\frac{1}{n} (\frac{{\cal R}_{11}'}{n{\cal R}_{11}})'
+ \frac{{\cal R}_{11}'^2}{n^2 {\cal R}_{11}^2})
-6(\frac{1}{n} (\frac{b'}{nb})' + \frac{b'^2}{n^2 b^2}) \nonumber \\
&&R_{kj} = (\frac{1}{n} (\frac{a'}{na})' + 3 \frac{a'^2}{n^2 a^2}
+ \frac{{\cal R}_{11}'a'}{n^2{\cal R}_{11}a}
+ 6 \frac{a'b'}{n^2ab}) \delta_{kj} \nonumber \\
&&R_{\varphi\varphi} = \frac{1}{n} (\frac{{\cal R}_{11}'}{n{\cal R}_{11}})'
+ \frac{{\cal R}_{11}'^2}{n^2 {\cal R}_{11}^2}
+ 3 \frac{{\cal R}_{11}'a'}{n^2{\cal R}_{11}a}
+ 6 \frac{{\cal R}_{11}'b'}{n^2{\cal R}_{11}b} \\
&&R_{MN} = {\cal R}_{MN} + (\frac{1}{n} (\frac{b'}{nb})'
+ 6 \frac{b'^2}{n^2 b^2} + \frac{{\cal R}_{11}'b'}{n^2{\cal R}_{11}b}
+ 3\frac{a'b'}{n^2ab}) \delta_{MN} \nonumber
\ea
Recalling that ${\cal R}_{MN} = 0$ for Calabi-Yau spaces, we see that the
equations of motion do not discern any intrinsic properties
of the Calabi-Yau spaces. Now, the Ricci scalar is
\ba
\label{riccis}
R &=& \frac{2}{n} (\frac{{\cal R}_{11}'}{n{\cal R}_{11}}
+ 3 \frac{a'}{na} + 6 \frac{b'}{nb})'
+ 2 \frac{{\cal R}_{11}'^2}{n^2{\cal R}_{11}^2}
+ 12 \frac{a'^2}{n^2a^2} \nonumber \\
&+& 42 \frac{b'^2}{n^2b^2}
+ 12 \frac{b'{\cal R}_{11}'}{n^2b{\cal R}_{11}}
+ 6 \frac{a'{\cal R}_{11}'}{n^2a{\cal R}_{11}} + 36 \frac{a'b'}{n^2a'b'}
\ea
It contains a term of second order in derivatives. We
will eliminate it from the effective reduced Lagrangian by a partial
integration, and omission of the ensuing boundary term, since we are
only interested in the bulk equations of motion here. The gravitational
Lagrangian is
${\cal L} = \sqrt{g} R = \sqrt{G_{CY}} n {\cal R}_{11} a^3 b^6
R = \sqrt{G_{CY}} L$. The action (\ref{act}) then is
\ba
\label{redact1}
S &=& \int d^{11}x \sqrt{g} R = \int dt d^3 \vec x d^6 y d\varphi
n {\cal R}_{11} a^3 b^6\sqrt{G_{CY}} R \nonumber \\
&=& \mu_R V \int dt L =
\mu_R V \int dt n {\cal R}_{11} a^3 b^6 R
\ea
where $\mu_R$ contains a finite renormalization of the
mass scale $M_{11}$ by the volume of the
Calabi-Yau $3$-fold ${\cal V}_6$ and the period $2\pi$ of $\varphi$.
$V$ is the comoving volume of the 3-space ${\cal M}_3^{k=0}$.

In addition to the gravitational part
of the action, the gauge terms give\footnote{We would like to thank
J.E. Lidsey for pointing out a missing sign in the earlier version
in terms proportional to $Q^2$.}
\be
\label{pot}
F^2_{\mu\nu\lambda\sigma} = - \frac{24}{a^6}
(\frac{{\cal F}_1^2}{n^2} - \frac{{\cal F}_2^2}{{\cal R}_{11}^2})
\ee
with
\be
\label{potf}
{\cal F}_1^2 = P^2 \frac{a^6}{b^{12} {\cal R}_{11}^2}
~~~~~~~ {\cal F}_2^2 = Q^2
\ee
When we substitute $-(1/48) F^2_{\mu\nu\lambda\sigma}$
into the action along with the explicit form of the Ricci scalar
(\ref{riccis}), we find that after omitting a boundary term, and using
$I = S/(\mu_R V)$, the action becomes
\ba
\label{redactfin}
I &=& \int \frac{dt}{n} \Bigl\{ \frac{{\cal R}_{11} b^6}{2a^3} {\cal F}_1^2
- \frac{n^2 b^6}{2{\cal R}_{11} a^3} {\cal F}_2^2 \nonumber \\
&&-{\cal R}_{11} a^3 b^6 \Big(6 \frac{a'^2}{a^2} + 30 \frac{b'^2}{b^2} +
6 \frac{a'{\cal R}_{11}'}{a{\cal R}_{11}}
+ 12 \frac{{\cal R}_{11}'b'}{{\cal R}_{11}b} + 36 \frac{a'b'}{ab}
\Bigr) \Bigr\}
\ea

To find the equations of motion for the remaining gravitational
degrees of freedom, we first vary this action with respect
to $n,a,b,{\cal R}_{11}$ and then insert the expressions for ${\cal F}$
in (\ref{potf}) - i.e. we treat ${\cal F}$ as a constant
under variations. This reproduces the correct equations
of motion, as is easy to check. Choosing the gauge $n=1$, and
introducing the
mini-superspace ``particle coordinates"
$\alpha = \ln(a)$, $\beta = \ln(b)$
and $\gamma = \ln({\cal R}_{11})$,
we obtain the following equations of motion:
\ba
\label{parti1}
&&6 \alpha'^2 +
30 \beta'^2 + 6 \alpha'\gamma' + 12 \beta'\gamma'
+ 36 \alpha'\beta' = \frac{P^2}{2} e^{-2\gamma - 12 \beta} + \frac{Q^2}{2}
e^{-2\gamma-6 \alpha}
\nonumber \\
&&\alpha'' + 3 \alpha'^2 + 6 \alpha'\beta'
+ \alpha'\gamma'= -\frac{P^2}{3} e^{-2\gamma - 12\beta}
+\frac{Q^2}{3} e^{-2\gamma - 6\alpha} \nonumber \\
&&\beta'' +
6 \beta'^2 + 3 \alpha'\beta' + \beta'\gamma'
= \frac{P^2}{6} e^{-2\gamma - 12\beta} -
\frac{Q^2}{6} e^{-2\gamma - 6\alpha} \\
&&\gamma'' + \gamma'^2
+ 3 \alpha'\gamma' + 6 \beta'\gamma' =
\frac{P^2}{6} e^{-2\gamma - 12\beta}
+\frac{Q^2}{3} e^{-2\gamma - 6\alpha} \nonumber  \nonumber
\ea
These equations resemble the equations of motion of a
mechanical system evolving with friction (the terms
bilinear in first derivatives).
The constraint equation can be
thought of as a generalized energy integral.
To make the mechanical analogy for (\ref{parti1}) more precise,
we will introduce a new gauge below, which will
remove the friction terms.

At this point, however, it is illustrative to review
the $11D$ Kasner solutions, defined by setting $P=Q=0$.
The Kasner ans\"atz
corresponds to choosing $\alpha' = \alpha_0/t$,
$\beta' = \beta_0/t$ and $\gamma' = \gamma_0/t$. So
the equations (\ref{parti1}) give
\ba
\label{kasner1}
&&\alpha_0^2 + 5 \beta_0^2 +\alpha_0 \gamma_0 +
2 \beta_0 \gamma_0 + 6 \alpha_0 \beta_0 = 0 \nonumber \\
&&\alpha_0 = 3 \alpha_0^2 + 6 \alpha_0\beta_0
+ \alpha_0\gamma_0 \nonumber \\
&&\beta_0 = 6 \beta_0^2 + 3 \alpha_0\beta_0
+ \beta_0\gamma_0 \\
&&\gamma_0 = \gamma_0^2
+ 3 \alpha_0\gamma_0 + 6 \beta_0\gamma_0 \nonumber
\ea
The solutions come in several varieties:
(1) if none of the $\alpha_0$, $\beta_0$
and $\gamma_0$ are zero, they must satisfy
$3 \alpha_0 + 6 \beta_0 + \gamma_0 = 1$ (as one can
easily see from the latter three equations, which
degenerate to this single equation) and
$3 \alpha_0^2 + 6 \beta_0^2 + \gamma_0^2 = 1$
(which arises after taking the square the equation
above and then subtracting from it the first equation
from (\ref{kasner1}) (2) a few degenerate cases, where
the possibilities for $(\alpha_0,~\beta_0,~\gamma_0)$ are
$(i)~(0,0,1)$, $(ii)~(1/2,0,-1/2)$, $(iii)~(-1/3,1/3,0)$,
$(iv)~(5/9,-1/9,0)$ and $(v)~(0,2/7,-5/7)$. In fact, case $(i)$
is locally just the 11D flat space in Milne coordinates, as can be
seen by applying a simple coordinate transformation. To make the
contact with string models we have discussed earlier,
in particular with the action (\ref{act10s}), we note
that solutions $(iii)$ and $(iv)$ correspond to
10D modular string cosmology solutions with the constant dilaton field,
while $(ii)$ and $(v)$ can be understood as
particular solutions with both rolling dilaton
and rolling moduli fields. The generic Bianchi models where all the
scale factors depend on time also correspond to string cosmologies
with rolling dilaton and moduli fields. However, the two cases
where $2\beta_0+\gamma_0 = 0$, and $0 \ne \alpha_0 \ne \beta_0 \ne 0$
produce precisely the metric-dilaton string solutions (\ref{metdil}),
as can be immediately verified by following the procedure
outlined above, in the discussion leading to
(\ref{act10s}). Namely, first dimensionally reduce from 11D to 10D type
IIa string theory and then down to 4D assuming that the internal six
dimensions span an isotropic six-torus.

Finally, perhaps the most curious interpretations of these
solutions arise in the following way.
There may exist reduction procedures which
map $11D$ solutions to lower dimensional stringy ones, and in
particular $4D$ solutions, but in a way which involves a phase
of M-theory not belonging to known string theories. For example,
if we take the apparently trivial case $(i)$ or the curved
case $(ii)$, we can map them both onto the same metric-dilaton
solutions (\ref{metdil}). The solutions
however, acquire the guise of string theory only at the very end -
in both cases we first perform the dimensional reduction on the Calabi-Yau
space, producing a simple Einstein theory in five dimensions
as a result. To check that this is a consistent truncation of the $11D$
theory one only need recall that the equations of motion
for the Calabi-Yau scale factor are homogeneous (see e.g. (\ref{parti1})).
Then, in a manner similar to that discussed by
Behrndt and F\"orste in \cite{mology}, we go one dimension
down to four, obtaining an action of a scalar-tensor theory of
gravity in a Brans-Dicke frame:
\be
\label{stbd}
S = \int d^4x \sqrt{\hat g} {\cal R}_{11} \hat R
\ee
The hats denote the Brans-Dicke frame quantities.
In this frame, our solutions $(i)$ and $(ii)$ are
${\cal R}_{11} = t$, $d\hat s^2 = - dt^2 + d\vec x^2$ and
${\cal R}_{11} = 1/\sqrt{t}$, $d\hat s^2 = - dt^2 + td\vec x^2$.
To see that both of these solutions are conformal to
the solution (\ref{metdil}), we perform another
conformal transformation, to the string frame. For both of these
two cases, the conformal transformation and the field redefinition
of the scalar field look formally the same. They are
$g_{\mu\nu} = {\cal R}^{1\pm \sqrt{3}}_{11} \hat g_{\mu\nu}$
and $\phi = \pm (\sqrt{3}/2) \ln {\cal R}_{11}$ (Note that these
field redefinitions are slightly different from those used to obtain eq.
(\ref{act10s}) because we are now reducing from 5D to 4D as
opposed to from 11D to 10D). Since in the two cases  the radius of the eleventh
dimensions depends on time differently, we need to carry out coordinate
transformations separately. In case $(i)$, we thus find the string-frame
comoving time to be
$\tau \sim t^{(3\pm \sqrt{3})/2}$,
while for case $(ii)$ the transformation is
$\tau \sim t^{(3\pm \sqrt{3})/4}$. When we substitute the
field redefinitions and these coordinate transformations into
the solutions, both cases lead precisely to (\ref{metdil}).
As one can now see, this solution (\ref{metdil}) is highly degenerate
from the point of view of M-theory as the same solution in string  theory
can be obtained from several different solutions in
M-theory!

Reductions of this type have not been
given much attention before, since they employ a dimensional descent
inside M-theory, making contact with strings only at the very end.
However, they are none the less interesting, since by being solutions
of the 11D supergravity equations of motion they certainly belong
to the general phase space of M-theory. An interesting
feature of the case $(i)$ discussed above is that
it offers a reinterpretation
of the Pre-Big-Bang curvature singularity in (\ref{metdil}) as a
Rindler horizon in 11D, in a way very similar to what
has been discussed in \cite{dtype}. However, since the horizon
involves the compact coordinate along the circle ${\cal S}^1$ of
(\ref{topology}), the singularity has not been completely
removed from the 11D geometry. Instead, the
periodicity of $\varphi$ implies that the Rindler wedges
of the 11D manifold contain closed time-like curves, and moreover
that the manifold is not Hausdorff, as observed in \cite{dtype}.
A difference between our example and those of \cite{dtype}
is that we use the $11^{\rm th}$ direction to define the
horizon, and hence lift the singularity, thus going outside of the
realm of string theory constructions. In this case, we regulate
the $4D$ string coupling by decompactifying directly into the
$11D$ supergravity phase,
rather than trying to stabilize the coupling within the string framework.
The benefit of this approach is the softening of the
singularity, namely the curvature singularity is absent, but the
space-time exhibits geodesic incompleteness. However, this still cannot be
taken as an example of graceful exit in Pre-Big-Bang. Such an extension
of the solution  involves an ascent to
$11D$ supergravity which does not belong to the original Pre-Big-Bang
scenario. Furthermore, at the transition, the function which corresponds
to the effective string coupling is really very small, as opposed to very
large - which is in  contrast to the generic situation in Pre-Big-Bang
scenario. The flow of the coupling in this example is opposite
to the flow of the coupling in various implementations of the
Pre-Big-Bang. In a sense, this example appears to accomplish one of the
goals of Pre-Big-Bang - singularity softening - while failing to
attain the other - pole inflation.

\section{\bf Cosmic Branes}

Returning to the general case with two charges, we will use
a more suitable gauge to examine solutions of
(\ref{parti1}). We note that
the terms which are bilinear in the first derivatives
of the fields $\alpha$, $\beta$ and $\gamma$
in (\ref{parti1}) are always
proportional to $ (3 \alpha' + 6 \beta' + \gamma')$.
In fact, this is the reason why the three equations
of motion in the Kasner
case degenerated to just one. This implies that we can gauge
away all such terms by using
a different time coordinate.
So let $dt = n d \tau$,
where $n = \exp(3 \alpha + 6 \beta + \gamma)$. Then we can
rewrite the equations of motion (\ref{parti1})
as (where the overdots
denote $\tau$ derivatives)
\ba
\label{parti2}
&&6\dot \alpha^2 +
30 \dot \beta^2 + 6 \dot \alpha \dot \gamma + 12 \dot \beta \dot \gamma
+ 36 \dot \alpha \dot \beta = \frac{P^2}{2} e^{6\alpha}
+\frac{Q^2}{2} e^{12\beta}  \nonumber \\
&&\ddot \alpha = -\frac{P^2}{3} e^{6\alpha}
+\frac{Q^2}{3} e^{12\beta} ~~~~~~~~~~~~~
\ddot \beta = \frac{P^2}{6} e^{6\alpha} -
\frac{Q^2}{6} e^{12\beta}  \\
&&~~~~~~~~~~~~~~~~~~~~~~\ddot \gamma =
\frac{P^2}{6} e^{6\alpha}
+\frac{Q^2}{3} e^{12\beta} \nonumber
\ea
These equations admit the mechanical analogy we have
indicated at the end of the previous section.
The constraint can be thought of as the
conservation of energy - it is just the Hamiltonian,
with the requirement that $E=0$.  If we take the derivative-dependent
terms in the constraint to denote the kinetic energy, and the
exponentials to be the potential, such that $H = T + W$,
we can define the Lagrangian as $L = T - W$.
One can then show that
the second order equations of motions follow from the variation
of this Lagrangian.
Similar equations of motion were considered recently in
\cite{form}.

Before continuing with the investigation of the
general solutions, we will first review
the cases when one of the charges is zero. These cases were considered in
\cite{Lu}, \cite{Luk}, \cite{clw}, and we include them here for
completeness. We will see that the two
possibilities lead to different subclasses of solutions. Let us begin
with the case $Q=0$. This case corresponds to the axionic
cosmology extended to 10D by the
addition of rolling moduli, which has been investigated recently
by Copeland, Lidsey and Wands \cite{clidw}.
The equations of motion reduce to
\ba
\label{parti2Q0}
&& 6\dot \alpha^2 +
30 \dot \beta^2 + 6 \dot \alpha \dot \gamma + 12 \dot \beta \dot \gamma
+ 36 \dot \alpha \dot \beta = \frac{P^2}{2} e^{6\alpha} \nonumber \\
&&\ddot \alpha = -\frac{P^2}{3} e^{6\alpha} ~~~~~~~~~
\ddot \beta = \frac{P^2}{6} e^{6\alpha} ~~~~~~~~~
\ddot \gamma =
\frac{P^2}{6} e^{6\alpha}
\ea
and it is clear that they are easily integrable - all we need to
do is solve the $\alpha$ equation, and the rest reduce to simple double
integrals. The $\alpha$ equation is the familiar Liouville equation.
It can be integrated once to give the first integral
\be
\label{Liuint}
\dot \alpha^2 = \theta^2_0 - \frac{P^2}{9} e^{6 \alpha}
\ee
where $\theta^2_0$ is a positive integration constant (if it were
zero or negative, we would have found $\dot \alpha=0$ and so $P=0$).
This separates variables, and we can rewrite it as the integral
\be
\label{Liusol}
\int \frac{d(e^{-3\alpha})}{\sqrt{\theta^2_0 e^{-6\alpha} - P^2/9}}
= \mp 3 (\tau + \tau_0)
\ee
where $\tau_0$ is another integration constant. We can set it to zero
by a time translation. After all the integrations, we find
\ba
\label{Liusol1}
&&e^{-3\alpha} = |\frac{P}{3\theta_0}| \cosh 3\theta_0 \tau
~~~~~~~~~~
e^{6 \beta} = e^{6 \beta_0 + 6 \beta_1 \tau} \cosh 3
\theta_0 \tau \nonumber \\
&&~~~~~~~~~~~~~~~~~
e^{6 \gamma} = e^{6\gamma_0 + 6 \gamma_1 \tau} \cosh 3 \theta_0 \tau
\ea
The constraint relates the integration constants according to
$30\beta^2_1 + 12 \beta_1 \gamma_1 = 9 \theta^2_0/2$. These solutions
are valid for both $\tau >0$ and $\tau<0$, and $\tau = 0$ is
the singularity. Note that we can take the limit when $\gamma = \beta$,
by adjusting the integration constants, in which case this reduces
to the solutions discussed very recently by Lukas and Ovrut \cite{luov1}.

Consider now the $P=0$ cases. Given the structure of the equations
of motion, one may expect that there
should be a kind of duality correspondence between the $P=0$ and $Q=0$
cases. Indeed, the equations of motion are
\ba
\label{parti2P0}
&& 6\dot \alpha^2 +
30 \dot \beta^2 + 6 \dot \alpha \dot \gamma + 12 \dot \beta \dot \gamma
+ 36 \dot \alpha \dot \beta = \frac{Q^2}{2} e^{12\beta}  \nonumber \\
&&\ddot \alpha = \frac{Q^2}{3} e^{12\beta} ~~~~~~~~~~
\ddot \beta = -\frac{Q^2}{6} e^{12\beta}     ~~~~~~~~~~
\ddot \gamma = \frac{Q^2}{3} e^{12\beta}
\ea
Note that now we need to solve the $\beta$ equation, which again
is a Liouville equation, and the rest follows easily. First, we again
get the integral of motion,
\be
\label{firintq}
\dot \beta^2 = \theta_1 - \frac{Q^2}{36} e^{12 \beta}
\ee
where $\theta_1$ is an integration constant, which must be a positive
number. Then the solutions are
\ba
\label{solpos}
&&e^{-6\beta} = |\frac{Q}{6\sqrt{\theta_1}}| \cosh (6 \sqrt{\theta_1} \tau)
~~~~~~~~~~~~~~
e^{3\alpha} = e^{3\alpha_0 + 3 \alpha_1 \tau} \cosh(6 \sqrt{\theta_1} \tau)
\nonumber \\
&&~~~~~~~~~~~~~~~~~~
e^{3 \gamma} =  e^{3\gamma_0 + 3\gamma_1 \tau} \cosh(6 \sqrt{\theta_1} \tau)
\ea
and the constraint gives $\alpha_1^2 + \alpha_1 \gamma_1 = 3 \theta_1$.
This case indeed represents the dual of the case $Q=0$ considered before.
Hence, this sub-family is equivalent to the sub-family $Q=0$,
apart from the differences induced by the change in the numerical
value of the coupling in the exponentials. 

What can we do with the equations (\ref{parti1})
when both charges are nonzero?
Consider the equations for
$\ddot \alpha$ and $\ddot \beta$. If we look at their linear
combinations, we can see that $\ddot \alpha + 2 \ddot \beta =0$.
This gives us one integral of motion:
$\alpha + 2\beta = c (\tau - \tau_0)$.
Here $c$ and $\tau_0$ are integration constants.
We can choose $\tau_0 = 0$, by a time translation.
This allows us to rewrite the equations (\ref{parti2})
in terms of only two variables: $\alpha$ and $\gamma$. We find
\ba
\label{parti3}
&&~~~~~~~~~~~~15 c^2 -9\dot \alpha^2 + 6 c \dot \alpha + 12 c \dot \gamma
= P^2 e^{6\alpha}
+ Q^2 e^{6 c\tau - 6 \alpha}  \nonumber \\
&&\ddot \alpha = -\frac{P^2}{3} e^{6\alpha}
+\frac{Q^2}{3} e^{6 c\tau - 6 \alpha} ~~~~~~~~~~~~~~
\ddot \gamma =
\frac{P^2}{6} e^{6\alpha}
+\frac{Q^2}{3} e^{6 c\tau - 6\alpha}
\ea
These equations are not
easily integrable. In order to
obtain information about these cases, we have to resort
to numerical methods.

It should be clear that the initial sizes of the three subspaces
in (\ref{metbackground}) are not independent parameters on their
own. Rather, they combine with the charges $P$ and $Q$ and the
eleven-dimensional Planck mass (which enters by defining the time
scale of the evolution) to give the relevant parameters
for numerical integration. Hence we can simply set all of them to
one at the beginning and vary the two charges.
Here it is useful to begin by first classifying cosmological
solutions according to the relative signs of time derivatives
of the three scale factors in (\ref{metbackground}). According to
this classification there are {\it a priori} eight possibilities
(since each initial ``scale velocity" can be either positive or
negative). However, by time reversal we can correlate
the subclasses, and arrive at the conclusion that we only need to consider
four types of initial conditions, which we denote by ordered triplets
$({\it sgn}(\dot a_0), {\it sgn}(\dot b_0), {\it sgn}(\dot {\cal R}_{11~0})$:
$(+,+,+)$, $(+,+,-)$, $(+,-,+)$ and $(-,+,+)$.
The solutions however turn out
to be connected further by dynamics, as can be seen in the figures.
The cases $(+,-,+)$ and $(+,+,+)$ evolve into cases $(-,+,+)$,
in a way effectively similar to the generic idea of branch-changing
(which in this case does not correspond
to the exit, since all the solutions where ``branch-changing" occurs
posses singularities in the future).
This however does not mean that the generic behaviour of a cosmology
with two charges can be completely reduced down to only two cases.
Given the signs of initial values of ``scale velocities", we can study
different types of evolution depending on the ratios of the derivatives.
We present several typical cases
where we show the scale factors of the 3-space (a),
Calabi-Yau 3-fold (b) and
the circle $S^1$ (c). In Figure 1, we show a $(+,+,-)$ case, where
the circle initially shrinks from infinite size 
to a minimum size, when the contraction bounces and the circle
subsequently begins to expand, converging to a linear function
in time. This means that the coupling is initially flowing from the strong to
weak regime, where it bounces and returns towards the strong coupling,
but at a controllable rate. This is 
in contrast to the conventional Pre-Big-Bang solutions.
The 3-space begins with a zero radius and in the average 
continues to expand forever, undergoing selfsimilar 
osillations around the average expanding trajectory. These
phases consist of a stage of expansion, followed by 
a contraction and so on, ad infinitum. The average
expansion rate is however subluminal.
The Calabi-Yau space behaves very similarly to the 3-space, but
at a different scale, reflecting the fact that the equations
of motion are invariant under $a^3 \leftrightarrow b^6$, 
$P^2 \leftrightarrow Q^2$. Note that the spikes which appear 
in the figures are perfectly
smooth, when we zoom in at the appropriate scale.
The example in Figure
2 shows precisely the same behavior as that give in Figure 1, where
however we see that for appropriate choice of initial
point, this case may describe our $(-,+,+)$, $(+,+,+)$ and $(+,-,+)$
cases, which are all therefore equivalent by evolution. 
Finally, we give as an example in Figure 3 a case corresponding to
$(-,+,-)$, which represents a time inversion of the dynamics shown
in Figures 1 and 2.

Aside from disclosing a number of solutions,
our numerical investigation suggests that the solutions in
Eq. (\ref{metdil})  when viewed as cosmological backgrounds of type IIA
string theory and  M-theory are really rather special in that they have
infinitely  old inflating past branches. Most charged solutions
appear to have spatial sections which evolve out of a past spatial
singularity, much like the $k=1$ solutions described in
Eq. (\ref{curvsols}).  This is clearly the effect of the $4$-form
charges,  and therefore it seems that the efficiency of the Pre-Big-Bang
scenario to produce inflation is very limited from the point of
view of type IIA string theory, because of the fine tuning problems
discussed in \cite{tw}. In these cases, many of the solutions either
recollapse too soon or emerge out of an initial singularity. 

\section{Conclusion}

In this article, we have considered several aspects of application
of M-theory in cosmology. Our main aim has been to cast the known string 
cosmological backgrounds, as well as some of their more straightforward 
generalizations, as $11$-dimensional metric-$4$-form configurations.
This point of view is useful in order to view the large coupling limits
of string cosmologies, which is ill-defined in string perturbation 
theory, but can be completely understood in terms of the decompactification
of the $11^{\rm th}$ dimension. Further, this allows for an egalitarian
description of the string moduli fields, which arise due to compactification, 
and can be related by $U$-duality maps. In the process of this $11$-dimensional
reinterpretation of solutions, we have shown that in some simple cases the
cosmological curvature singularities can be moderated. In particular, 
some of the known scalar-field dominated cosmological solutions are equivalent
to a flat $11$-dimensional space-time, which however has pathological 
topology that could involve closed time-like curves in the maximal extension.
The acausal domain is separated from the physical sector of the space-time
by a horizon, which upon dimensional reduction produces coupling, and curvature
singularities.

Before closing, we would like to note some aspects
regarding the initial singularity in the context of M-theory.
As we have discussed earlier, one goal of the Pre-Big-Bang scenario is the
smooth transition from a $(+)$-branch solution (which evolves towards
singularities) to a $(-)$-branch solution (which evolves away from
singularities).  In \cite{kmo1}, a solution to the
gravitational equations of motion based on the Damour-Polyakov ans\"atz
showed such a smooth transition.  By an appropriate choice of the
corrected coupling function $B(\phi)$, a non-singular solution has
been found.
However, the solution was not without its peculiarities.  In the two
solutions presented in \cite{kmo1}, the evolution of the dilaton caused
the function $B(\phi)$ to pass through zero (in fact this is a general
requirement of any such solution as it must utilize an ``egg" with $B<0$
(see \cite{kmo1} for details), essentially indicating a signature
change in $10$ dimensions.
However, the ten-dimensional picture may be
misleading, because in the context of M-theory,
we should only look at the function
$B(\phi)$ as being related to the radius of the 11th dimension.
There, it appears that we could view the evolution 
as a process of decompactification, where the
eleventh dimension  blows up and then shrinks again. From
this point of view, we need not see
any change of signature at all.
In this way, the compact radius would ``bounce" at infinity.
At this moment, in support of this we can only offer an
analogy with
the bounce picture in Liouville field theory of non-critical
strings proposed by one of us some time ago \cite{kogan},
which was used later in the study of a non-equilibrium
temporal flow and a closed-time-like path formalism for non-critical
strings by \cite{emn}. Because the
Liouville field is ultimately connected with time in non-critical string
theory, it is very tempting to think about possible connections between
these two pictures of evolution in M-theory and non-critical string
theory. We hope to return to this issue in the future.

Finally, we should mention the possibility of relating cosmological
evolution in M-theory we have discussed here with renormalization
group flows. As we have seen, during cosmological expansion not only
is the spatial scale $a(t)$ evolving with time, but so is the coupling
constant $\lambda(t)$. This means that one can consider the evolution of
the coupling constant $\lambda$ not as a function of time, but as a
function of the scale factor $a$, $\lambda = \lambda(a)$, 
leading to $d \lambda(a)/d \ln a = \beta(\lambda)$. 
The $\beta$ function $\beta(\lambda)$ can be easily calculated
for any particular cosmological solution.
It is interesting to ask whether or not this ``RG'' flow 
could correspond to an
actual quantum RG flow in a ten-dimensional theory or even in a
four-dimensional one after compactification. For example, it would be very
interesting to find an example featuring
logarithmic behavior rather than the above
scaling, which could give
inflationary expansion, $a(t) \sim  \exp(c t)$, in the
lower-dimensional manifold. Unfortunately, we could
not find any M-theoretic cosmological solution with exponential inflation
in the physical
three-dimensional space and power-law evolution in the eleventh dimension.
However, this posibillity does not seem excluded at this point.
Indeed, solutions with such 
behavior would be of great interest, as they could be used for simultaneous
inflation with running coupling.
Hence, an affirmative answer to the question of existence of such solutions 
would be an extremely interesting result.
A conceptual difficulty to surpass here
is that  the cosmological evolution under consideration
is purely {\it  classical} by
definition - no quantization of $M$ theory was performed in our
analysis (simply because it does not exist  yet) - while the ordinary RG flow
is an entirely {\it quantum} phenomenon.
In a way, the situation is reminiscent
of the anomaly treatment in Wess-Zumino-Witten-Novikov $\sigma$-models,
where in the fermionic picture, the anomaly is quantum while in the bosonic
is appears purely classical. Perhaps a similar connection can be
established here. Indeed, the fact
that in the classical formulation of M-theory
one may have a flow of coupling constant
is quite interesting and certainly deserves further investigation.

\vspace{1cm}
{\bf Acknowledgements}

We would like to thank A. Linde, R. Kallosh,
J. Rahmfeld and G. Ross  for useful conversations. 
We would further like to extend our thanks to J.E. Lidsey
for pointing out a missing sign in terms which depend on $Q^2$
in the earlier version of the manuscript.
I.I.K. was supported
in part by PPARC,  Royal Society and Lockey foundation  travel grants
and K.A.O. was supported in
part by  DOE grant DE--FG02--94ER--40823. This paper was started when
N.K. and I.I.K  visited University of Minnesota in the summer 1997
and they are grateful for Department of Physics and Institute of
Theoretical Physics for hospitality and stimulating environment.

\newpage


{\bf Figure 1:} {The scale factors of the 3-space
(a), Calabi-Yau 3-fold (b) and
the circle $S^1$ (c). The 3-space and the Calabi-Yau space 
undergo repetitive stages of expansion and contraction, starting out of the 
initial singularity.  The circle initially shrinks, 
corresponding to the flow of the coupling from the strong to
weak regime, and reaches its minimum size after which it starts to
expand again, in an almost linear fashion. This is 
in contrast to the conventional Pre-Big-Bang solutions.}

{\bf Figure 2:} { In this example we see similar evolution
to that of Figure 1, but at different scales. }

{\bf Figure 3:} {The evolution in this case is a time-reversal
of the dynamics in the previous two figures. The 3-space
and the Calabi-Yau space undergo subsequent stages of expansion
and contraction, conracting more than expanding, during which time
the circle shrinks. This ends with a Big Crunch singularity for
both the 3-space and the Calabi-Yau, where the circle blow up.}

\eject
\newpage
\begin{figure}[t]
\vspace{-1.5truecm}
\epsfysize=3.0truein
\epsfbox{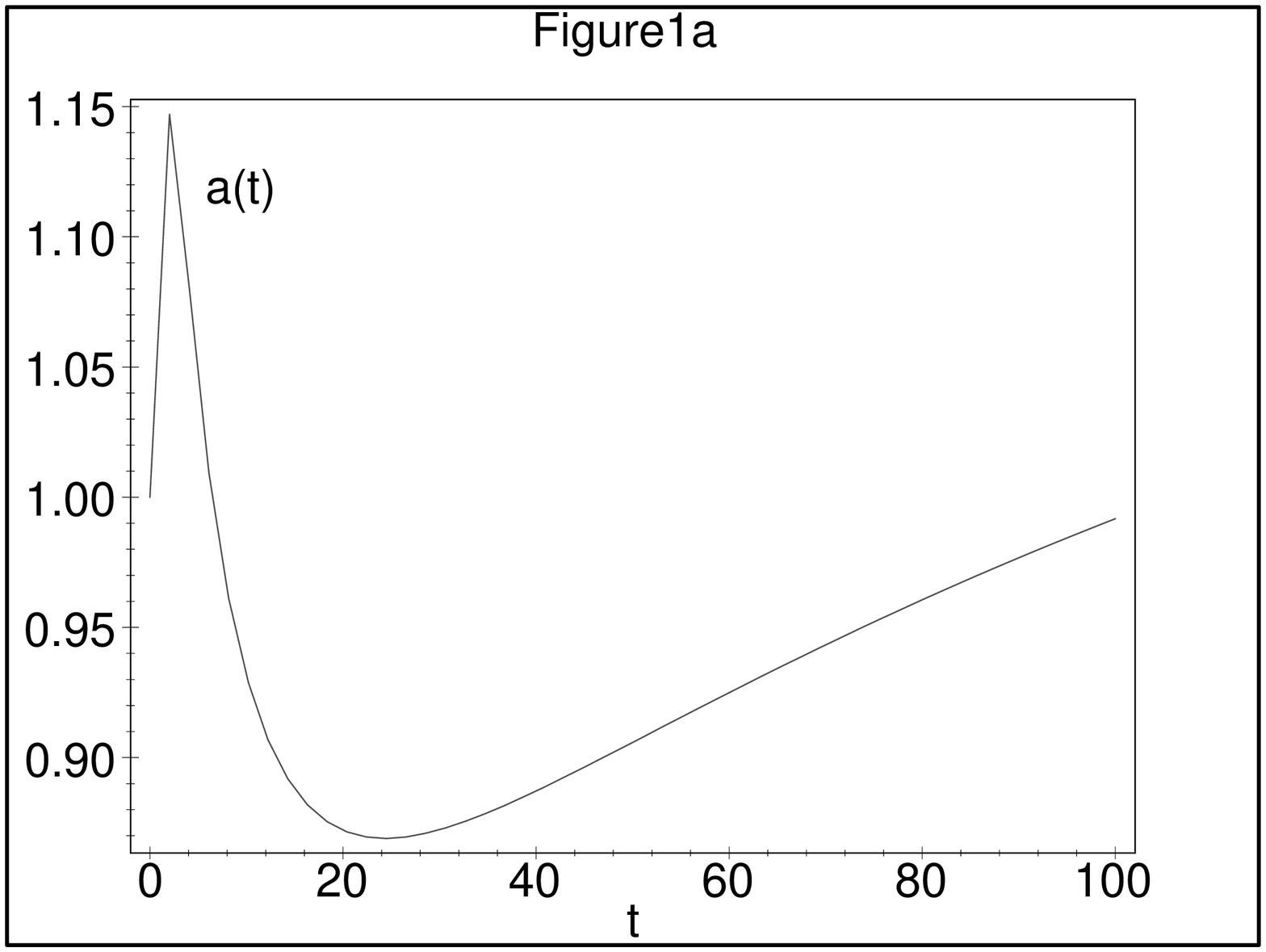}
\end{figure}

\begin{figure}[t]
\vspace{-1truecm}
\epsfysize=3.0truein
\epsfbox{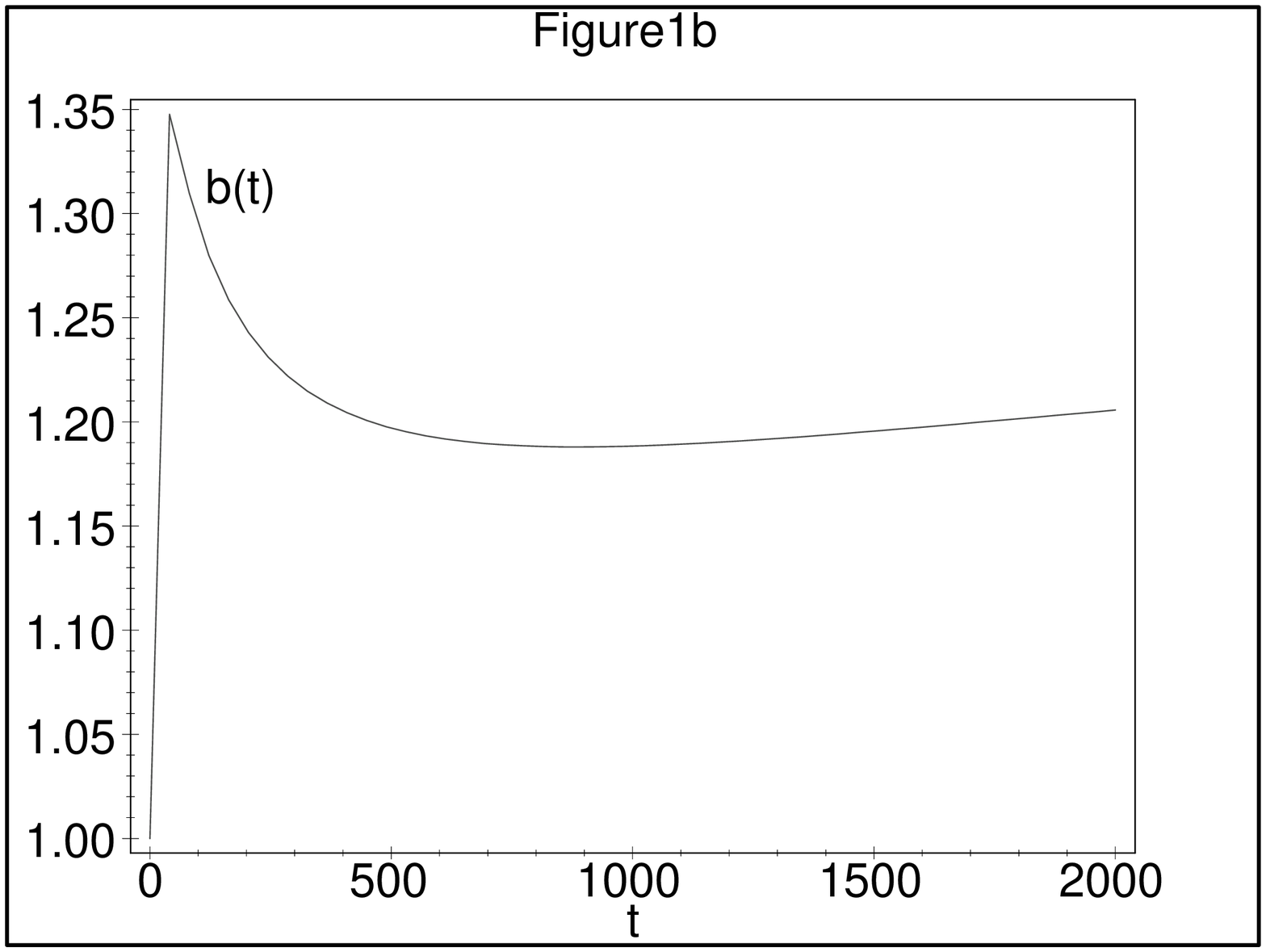}
\end{figure}

\begin{figure}[b]
\vspace{-1truecm}
\epsfysize=3.0truein
\epsfbox{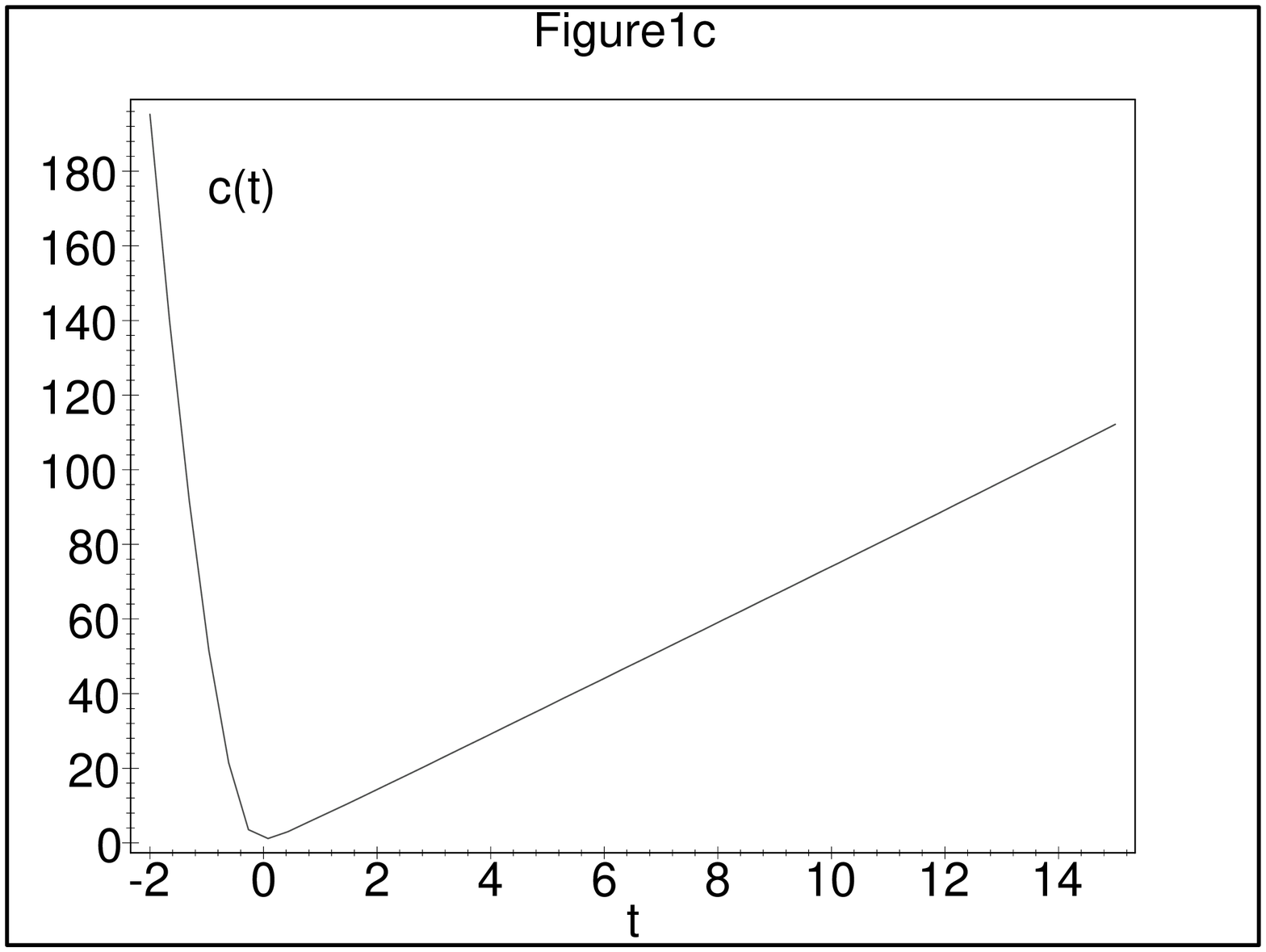}
\end{figure}

\begin{figure}[t]
\vspace{-1.5truecm}
\epsfysize=3.0truein
\epsfbox{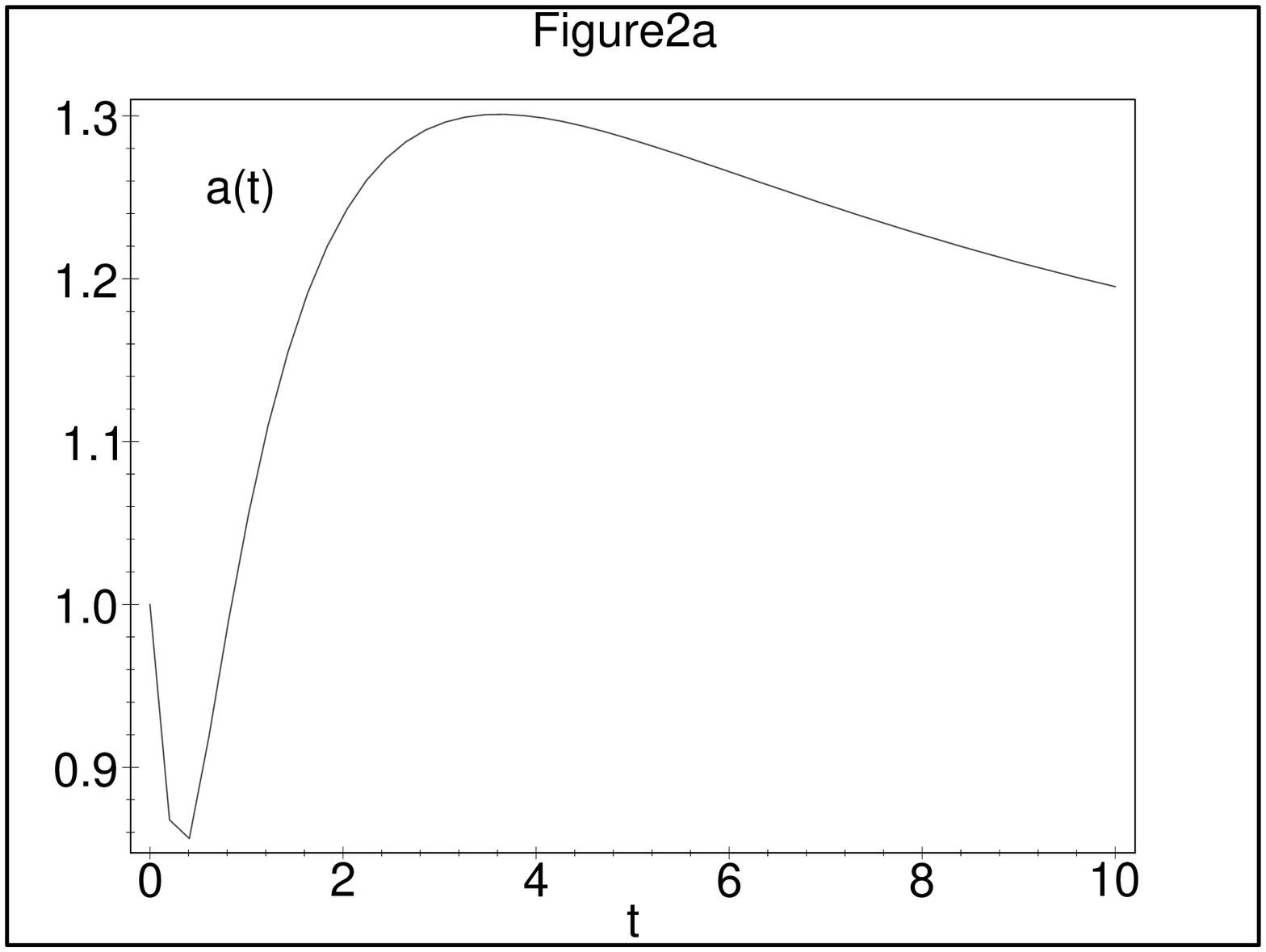}
\end{figure}

\begin{figure}[t]
\vspace{-1truecm}
\epsfysize=3.0truein
\epsfbox{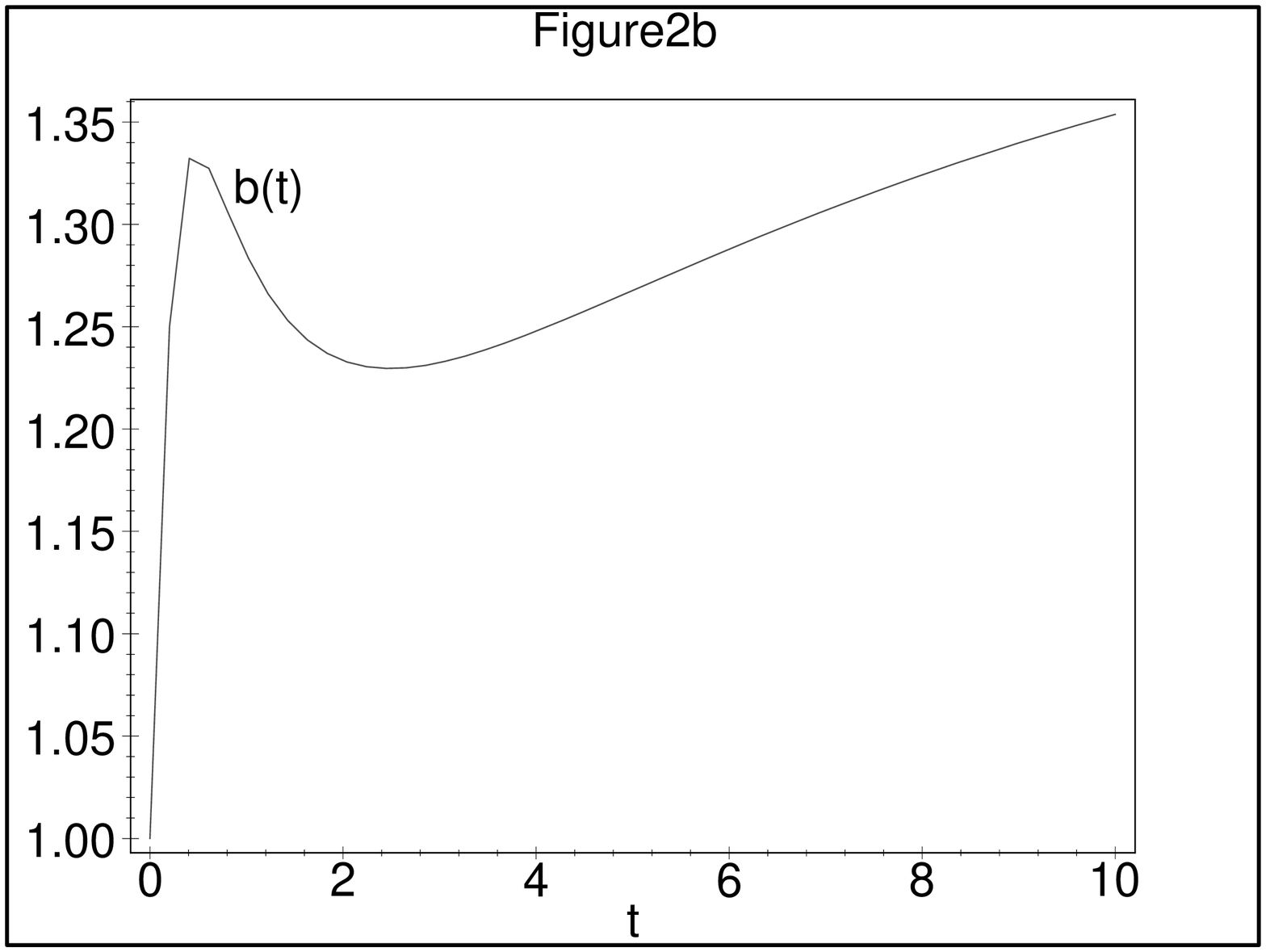}
\end{figure}

\begin{figure}[b]
\vspace{-1truecm}
\epsfysize=3.0truein
\epsfbox{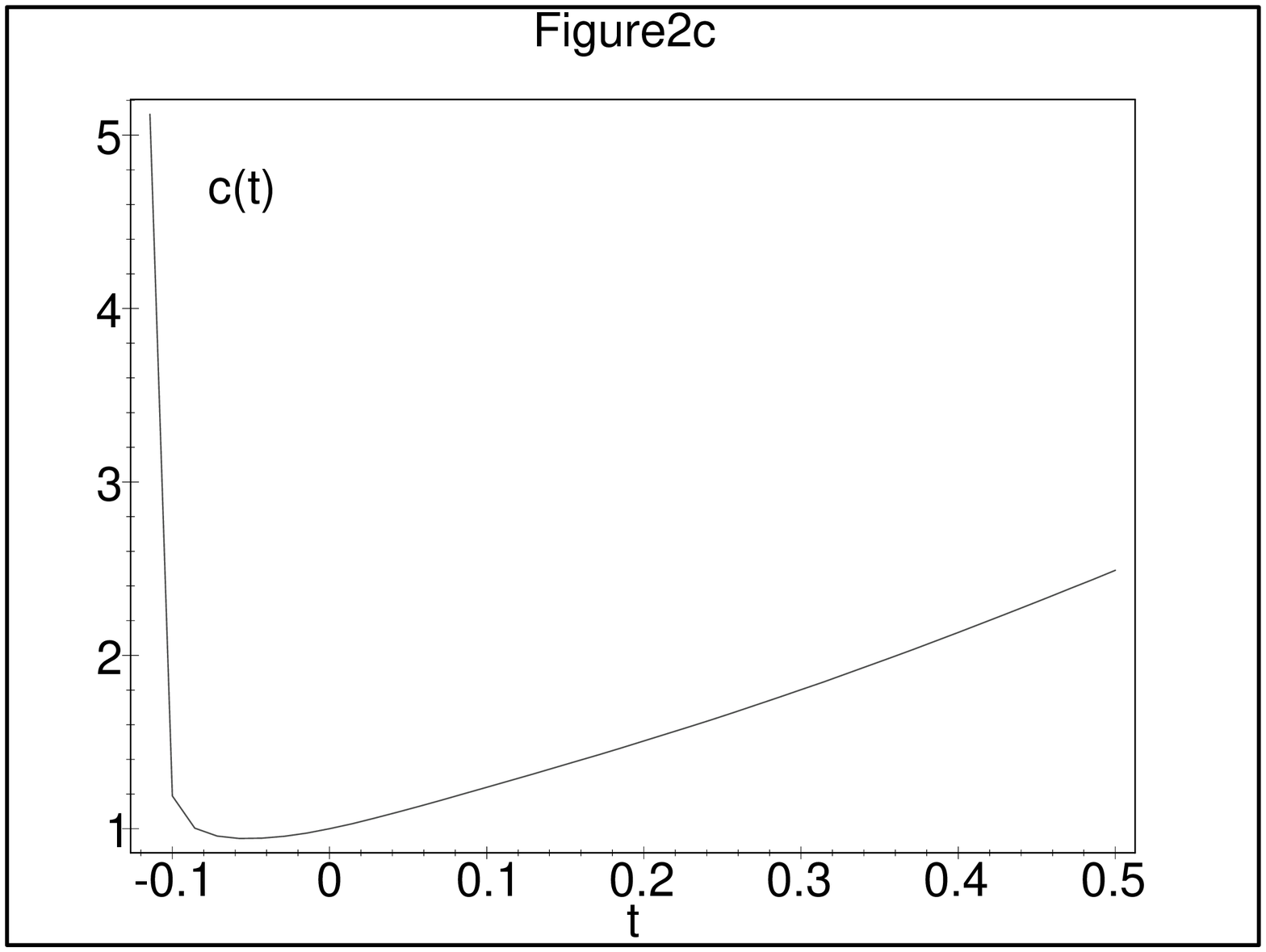}
\end{figure}

\begin{figure}[t]
\vspace{-1.5truecm}
\epsfysize=3.0truein
\epsfbox{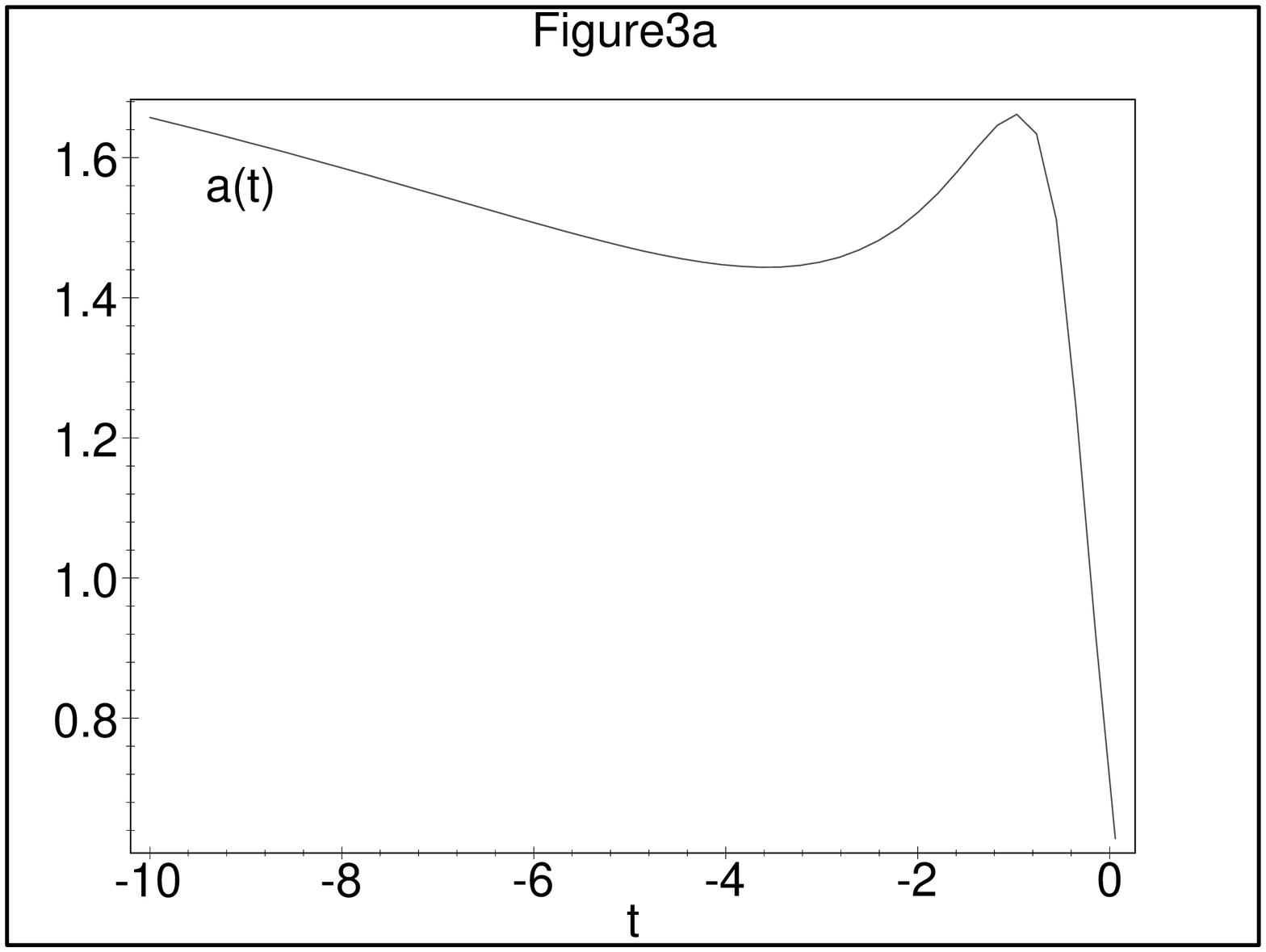}
\end{figure}

\begin{figure}[t]
\vspace{-1truecm}
\epsfysize=3.0truein
\epsfbox{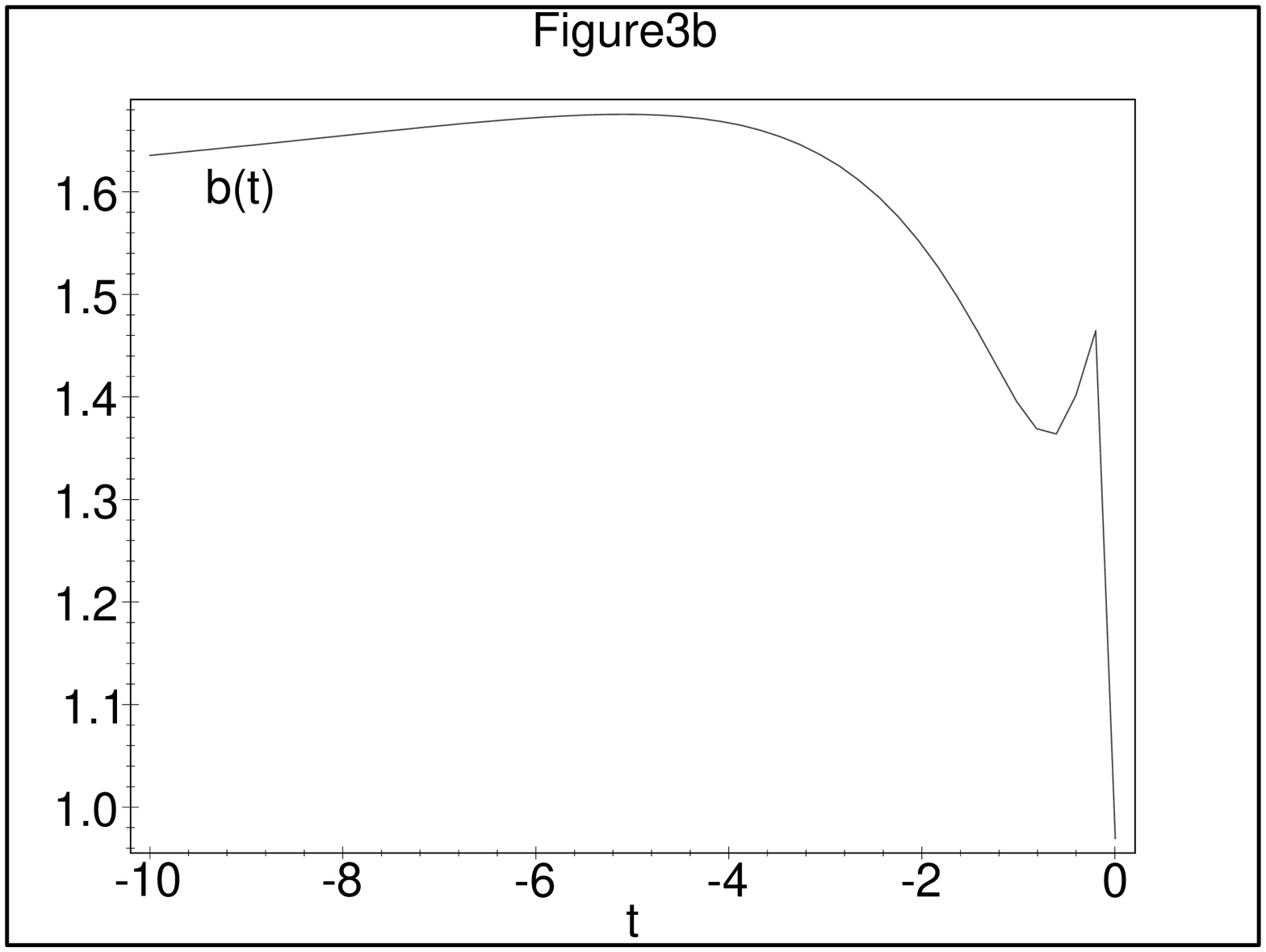}
\end{figure}

\begin{figure}[b]
\vspace{-1truecm}
\epsfysize=3.0truein
\epsfbox{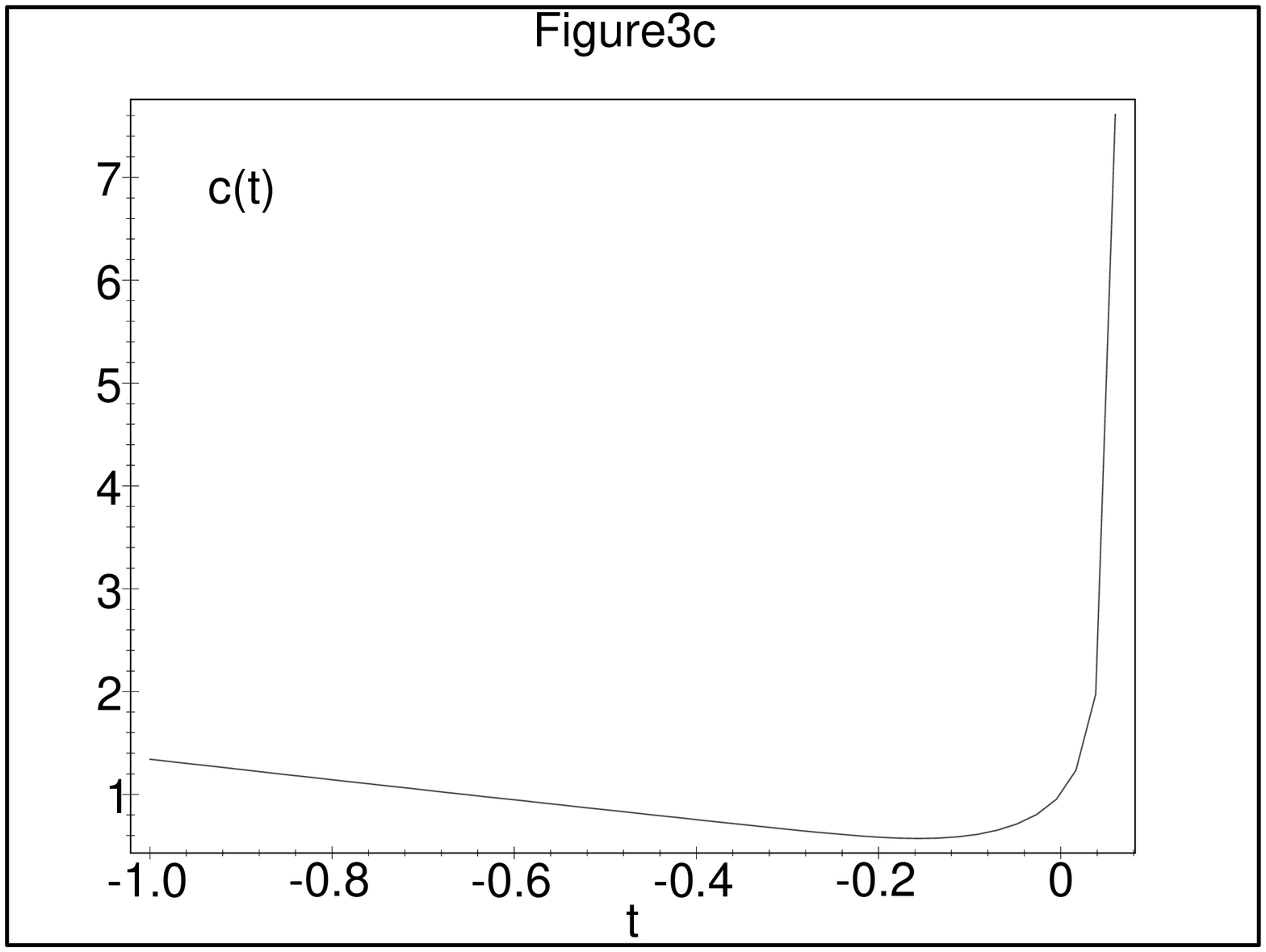}
\end{figure}

\end{document}